\pdfoutput=1
\documentclass[a4paper,12pt]{article}
\usepackage{amsfonts}
\usepackage{mathrsfs}
\usepackage{amsmath}
\usepackage{amssymb}
\usepackage[medium]{titlesec}
\usepackage{bm}
\usepackage{cite}
\usepackage[normalem]{ulem}
\usepackage{extarrows}
\usepackage{slashed}
\usepackage{isodateo}
\usepackage{graphicx}
\usepackage{xcolor}
\usepackage[bookmarksnumbered=true,bookmarksopen=true]{hyperref}
 \hypersetup{colorlinks,%
             linkcolor=[rgb]{0,0.3,0.6}, %
             citecolor=[rgb]{0,0.3,0.6}, %
             urlcolor=[rgb]{0,0.3,0.6}}
\usepackage[hmargin=.7in,vmargin=1.1in]{geometry}
\usepackage{indentfirst}
\newcommand{\FR}[2]{\displaystyle\frac{\,{#1}\,}{#2}}

\newcommand{\n}{\nonumber}

\graphicspath{{fig/}}

\def\bge{\begin{equation}}
\def\ede{\end{equation}}
\def\bga{\begin{aligned}}
\def\eda{\end{aligned}}
\def\bgp{\begin{pmatrix}}
\def\edp{\end{pmatrix}}
\def\bgs{\begin{subequations}}
\def\eds{\end{subequations}}
\newcommand{\order}[1]{\mathcal{O}({#1})}
\def\di{{\mathrm{d}}}
\def\D{{\mathrm{D}}}

\def\mb{\mathbf}

\def\pd{\partial}
\def\ld{{\mathscr{L}}}

\def\la{\langle}\def\ra{\rangle}
\def\sla{\slashed}

\def\to{\rightarrow}
\def\To{\Rightarrow}
\def\ii{\mathrm{i}}

\def\al{\alpha}
\def\be{\beta}
\def\ga{\gamma}
\def\de{\delta}
\def\ep{\epsilon}
\def\ka{\kappa}
\def\lam{\lambda}
\def\rh{\rho}
\def\si{\sigma}

\def\Mp{M_{\text{Pl}}}

\newcommand{\ob}[1]{\mkern 2mu \overline{\mkern -2mu #1 \mkern -2mu}\mkern 2mu}
\newcommand{\wt}[1]{\mkern 2mu \widetilde{\mkern -2mu #1 \mkern -2mu}\mkern 2mu}
\newcommand{\wh}[1]{\mkern 2mu \widehat{\mkern-2mu#1\mkern-2mu}\mkern 2mu}

\newcommand*{\vcenteredhbox}[1]{\begingroup
\setbox0=\hbox{#1}\parbox{\wd0}{\box0}\endgroup}

\begin{document} 

\title{ \Large\textbf{A Cosmological Higgs Collider}}

\author{
Shiyun Lu$^a$,~
Yi Wang$^{a}$,~
Zhong-Zhi Xianyu$^{b}$
\\[3mm]
\normalsize{$^a$~\emph{Department of Physics, The Hong Kong University of Science and Technology,}}\\
\normalsize{\emph{Clear Water Bay, Kowloon, Hong Kong, P.R.China}}\\
\normalsize{\emph{Jockey Club Institute for Advanced Study, The Hong Kong University of Science and Technology,}}\\
\normalsize{\emph{Clear Water Bay, Kowloon, Hong Kong, P.R.China}}\\
\normalsize{$^b$~\emph{Department of Physics, Harvard University, 17 Oxford St., Cambridge, MA 02138, USA}}
}

 \date{}
\maketitle

\vspace{2cm}

\begin{abstract}

The quantum fluctuations of the Higgs field during inflation could be a source of primordial density perturbations through Higgs-dependent inflaton decay. By measuring primordial non-Gaussianities, this so-called Higgs-modulated reheating scenario provides us a unique chance to probe Higgs interactions at extremely high energy scale, which we call the Cosmological Higgs Collider (CHC). We realize CHC in a simple scenario where the inflaton decays into Higgs-portal scalars, taking account of the decay of the Higgs fluctuation amplitude after inflation. We also calculate the CHC signals of Standard Model particles, namely their imprints in the squeezed bispectrum, which can be naturally large. The concept of CHC can be straightforwardly generalized to cosmological isocurvature colliders with other types of isocurvature perturbations.

\end{abstract}
 
\newpage

\section{Introduction}
\label{sec_intro}

The Higgs boson is a central focus of the current study of particle physics. It is the only fundamental scalar field experimentally discovered so far. It is also the only scalar particle in the particle Standard Model (SM), and is responsible for mass generation of SM particles via spontaneous electroweak symmetry breaking. On the other hand, the Higgs sector of SM suffers from a naturalness problem which hints at new physics beyond SM. The Higgs boson is also a unique portal to new physics. A careful study of Higgs properties may eventually lead us to a more fundamental theory beyond SM. (See e.g.~\cite{Tanabashi:2018oca} for a review.) 

The current strategy of studying the Higgs boson focuses almost exclusively on particle colliders. We have learned many properties about the Higgs boson from the LHC. Higgs physics is also among the most important physical targets of next-generation colliders operating at energies roughly within $\order{100\text{GeV}\sim 100\text{TeV}}$ \cite{CEPCStudyGroup:2018ghi,Bambade:2019fyw,Abada:2019zxq}. Given the importance of Higgs physics, it is desirable to explore alternative ways to study the Higgs boson, preferably at much higher energies. 

In this regard, the early universe provides us a unique chance. It is widely believed that an exponentially fast expansion of the universe, known as the cosmic inflation, has happened in the very early universe prior to the current phase of thermal big-bang expansion. (See \cite{Chen:2010xka, Wang:2013eqj, Baumann:2014nda} for reviews.) The energy scale of inflation, measured by the nearly-constant Hubble parameter $H$, can be as high as $\order{10^{13}}$GeV. Such a huge energy can trigger spontaneous particle productions through quantum fluctuations. The particle production can be effective for particles with mass up to $m\gtrsim H$, this covers essentially all SM particles, including the Higgs boson, and possibly many new particles beyond SM.

The idea of using the huge energy of inflation to probe heavy particles has been put forward in the context of the cosmological collider physics. See \cite{Chen:2009we, Chen:2009zp, Baumann:2011nk, Noumi:2012vr, Arkani-Hamed:2015bza,Lee:2016vti, Baumann:2017jvh} for previous works.  The essential idea is that, by couplings to the inflaton, the quantum fluctuations of heavy particles could leave characteristic imprints on $n$-point ($n\geq 3$) correlations of the inflaton field, known as the primordial non-Gaussianity. By measuring a particular ``channel'' --- the squeezed limit --- of 3-point function, one could extract the mass of the heavy particle as resonance peaks \cite{Chen:2009we, Chen:2009zp, Arkani-Hamed:2015bza} and its spin as angular distribution \cite{Arkani-Hamed:2015bza, Lee:2016vti, Baumann:2017jvh} of the correlation. Related ideas can turn the very early universe physics into classical \cite{Chen:2011zf, Chen:2011tu, Chen:2014cwa} and quantum \cite{Chen:2015lza, Chen:2018cgg} standard clocks to measure the expansion history of the very early universe.

Previous studies of cosmological collider assumed that the quantum fluctuations of the inflaton field seeded the large scale inhomogeneity and anisotropy. In this sense, we may say that the cosmological collider is an inflaton collider, which collects long-lived inflaton fluctuations. We study indirectly the dynamics of heavy fields by coupling them to the inflaton. This inflaton collider has been used to consider SM physics \cite{Chen:2016nrs,Chen:2016uwp,Chen:2016hrz,Kumar:2017ecc}, neutrino physics \cite{Chen:2018xck}, and other new physics \cite{Kumar:2018jxz}. In particular, it is shown that the mass of light particles will in general be lifted to the Hubble scale due to various types of mass corrections. (Light means lighter than the Hubble scale during inflation.) However, such corrections, together with the couplings to the inflaton, introduce lots of free parameters, making it difficult to extract and identify new physics from the observables. Worse still, it is in general difficult to get visibly large non-Gaussianities from heavy particles with generic couplings to the inflaton. Technically, the difficulty comes from the fact that visibly large non-Gaussianity prefers large inflaton-matter coupling, which could easily generate a large mass correction to the matter field and render it too heavy to be produced during inflation\footnote{There is an exception for this too-heavy-to-produce argument. The coupling between the inflaton and the heavy field may provide the heavy field a chemical potential. In the context of cosmological collider, an example of chemical potential is noted in \cite{Chen:2018xck}. With a chemical potential, very heavy particles can be produced making use of the inflaton kinetic energy.}.
 
Fortunately, there are alternatives that could be more interesting. In this paper we introduce the idea of a Cosmological Higgs Collider (CHC). The basic idea is that the large scale inhomogeneities and anisotropies could be seeded by the primordial fluctuations not of the inflaton field, but of the SM Higgs field.\footnote{Just to clarify that we are not identifying the Higgs field as the inflaton. We will assume a typical standard slow-roll inflation sector in this paper, except that we do not ask the inflaton field to generate all the primordial density fluctuations.} This is possible when the rate of the inflaton decay (and thus the efficiency of reheating) depends on the background value of the Higgs. In such cases, the fluctuations of the Higgs field will perturb the decay rate of the inflaton, and thus will perturb the expansion history of local universes. This is in line with the widely studied scenario of modulated reheating \cite{Dvali:2003em, Kofman:2003nx, Suyama:2007bg, Ichikawa:2008ne}, and we are simply identifying the Higgs boson as the light field modulating the inflaton decay. 

\begin{figure}[tbph]
\centering
\vcenteredhbox{\includegraphics[height=0.24\textwidth]{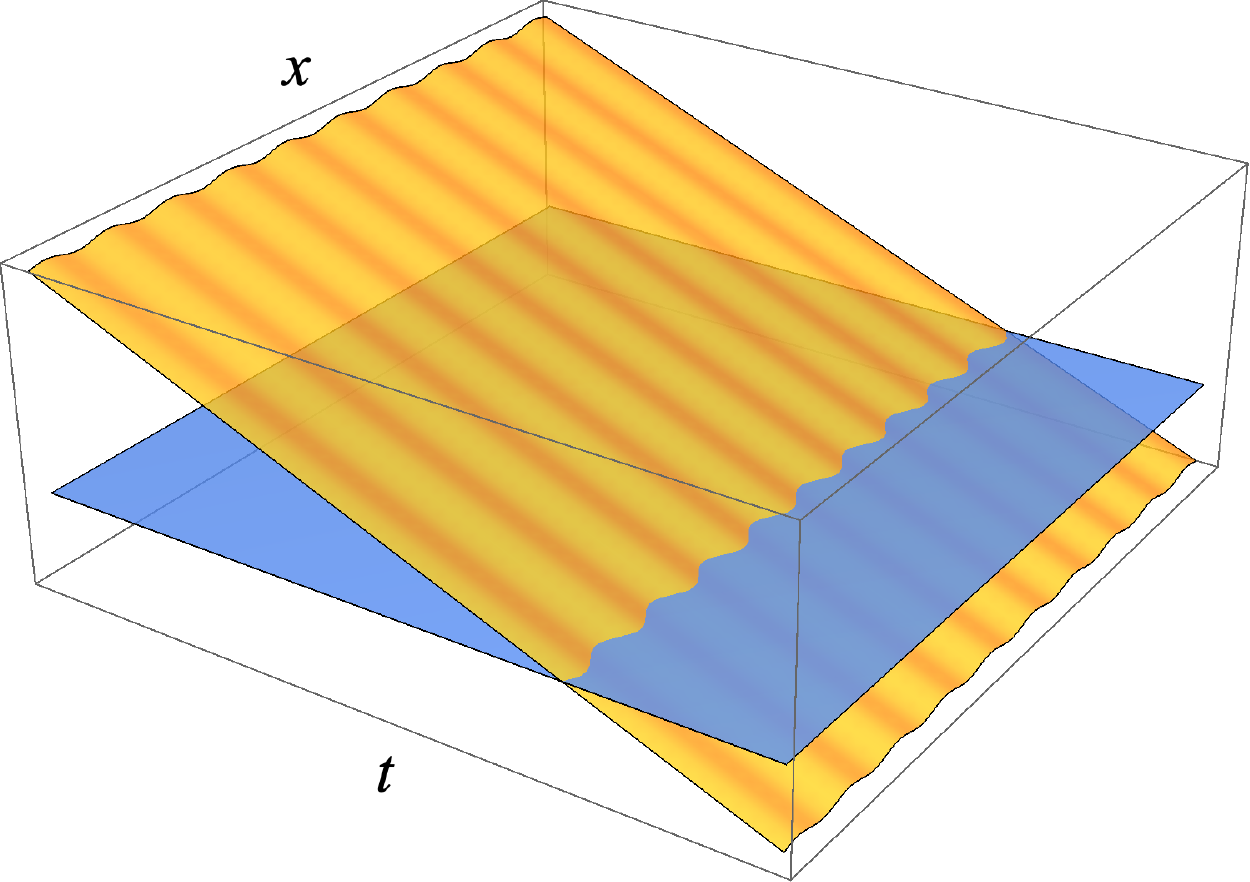}}
\hspace{10mm}
\vcenteredhbox{\includegraphics[height=0.24\textwidth]{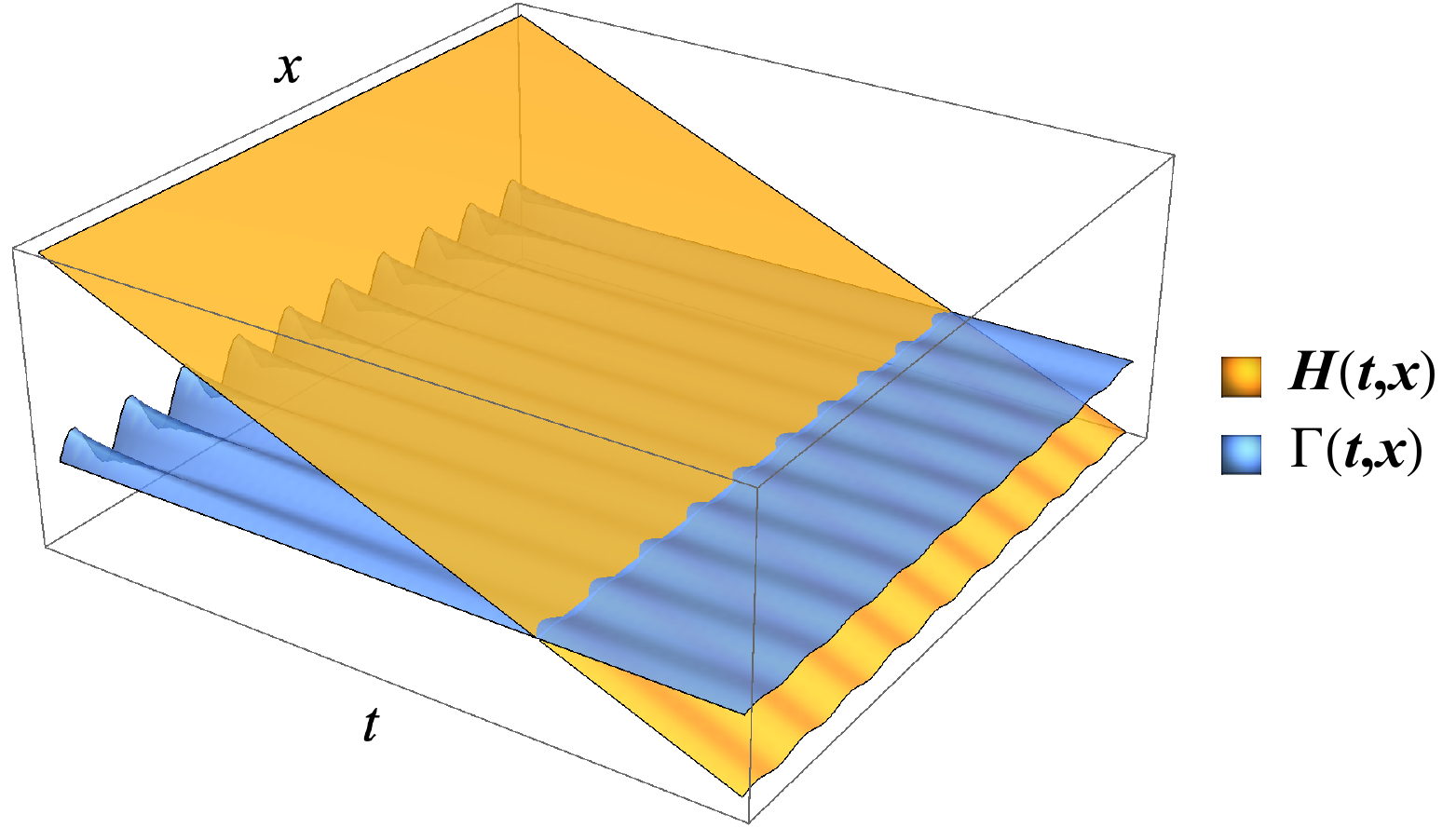}}
\caption{The ordinary inflation scenario versus modulated reheating, illustrating the Hubble scale (orange) and the decay rate of the inflaton (blue) as functions of time and space coordinates.}
\label{Fig_HGamma}
\end{figure}

To illustrate the difference between the modulated reheating scenario and the conventional inflation scenario, we show a sketch of the Hubble scale $H$ and the inflaton decay rate $\Gamma$ as functions of space position $x$ and time $t$ after inflation in Fig.\;\ref{Fig_HGamma}. In ordinary scenario (left panel), the local Hubble scale $H(t,x)$ (orange surface) was perturbed by the inflaton fluctuation during inflation (shown by the wiggly behavior in $x$ direction). After inflation, $H$ drops as a function of $t$. When it reaches the decay rate $\Gamma$ of the inflaton (blue surface), reheating is completed, and the spatial inhomogeneity of $H$ is translated to that of the temperature, which we can observe today.

In the modulated reheating scenario (right panel), the inflaton fluctuation is insignificant, so the Hubble scale $H(t,x)$ right after inflation is still quite homogeneous over space. However, when the inflaton decay rate is dependent on the background value of some light scalar fields (which we call the modulating fields), it can develop spatial inhomogeneities due to the primordial fluctuation of modulating fields. Then, during reheating, the spatial inhomogeneities in $\Gamma$ will be translated to that of the temperature, and also of the Hubble parameter. 

\begin{figure}[tbph]
\centering
\vcenteredhbox{\includegraphics[width=0.7\textwidth]{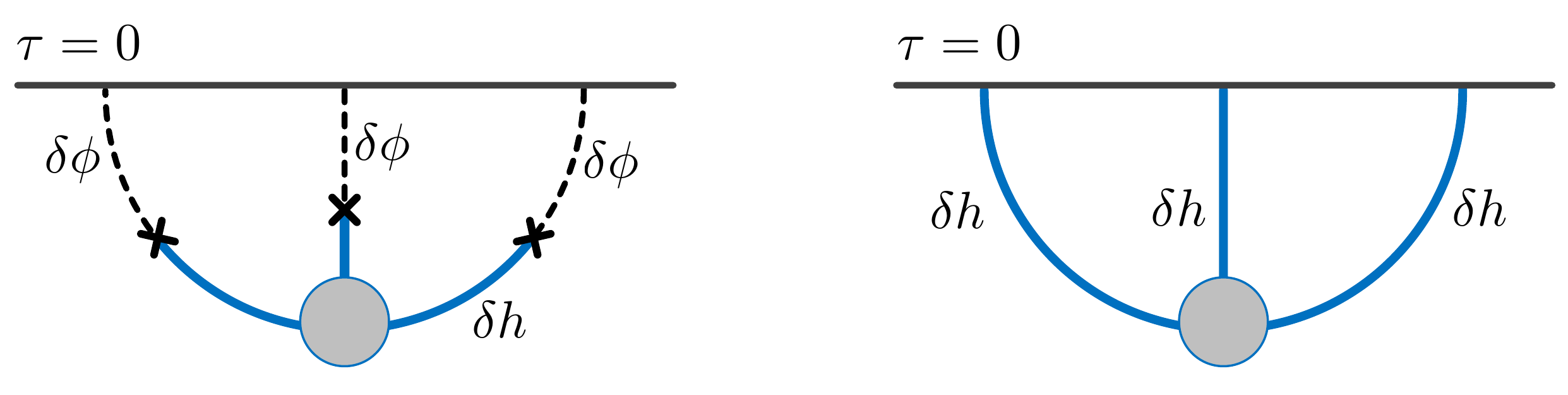}}
\caption{The original ``cosmological collider'' vs.\ the ``cosmological Higgs collider''.}
\label{Fig_CCvsCHC}
\end{figure}

When the SM Higgs boson acts as the modulating scalar, the large scale density fluctuations are directly related to the Higgs fluctuations. This allows us to study Higgs dynamics and interactions more directly during inflation. In other words, we are colliding Higgs directly in the cosmological collider and we can directly probe Higgs interactions and its interactions to other particles, be it SM or beyond SM, by measuring the primordial non-Gaussianity. In this sense we give the name of this scenario the ``Cosmological Higgs Collider'' (CHC). In Fig.\;\ref{Fig_CCvsCHC} we compare the ``particle collisions'' in the original cosmological collider and in the CHC. We see that the massive fields couple to Higgs directly in the CHC scenario (right panel). Their CHC signals are therefore directly related to their Higgs couplings rather than the inflaton couplings. Higgs couplings are usually large and thus easy to generate sizable non-Gaussianities. At the same time, the Higgs background, being order $H$ during inflation, does not give huge mass correction to the matter fields. Therefore, we see that the CHC is not only able to naturally generate sizable non-Gaussian signals, but also free from free-parameter ``pollutions'' from the unknown inflaton-matter couplings.

Previous studies have considered the scenario of Higgs-modulated reheating \cite{Choi:2012cp,DeSimone:2012gq,Cai:2013caa}. But it is not immediately obvious that the SM Higgs boson can be a modulating scalar. This is because the Higgs has a quartic potential which is not so flat. Therefore the fluctuation amplitude of the Higgs will decay after inflation. If the Higgs decay much faster than the inflaton, then at the time of reheating, the Higgs fluctuation would be too small to generate correct amount of density fluctuation as we measure today. This important effect of Higgs decay not only requires that the inflaton decays quickly enough, but also limits the form of Higgs dependence in the decay rate $\Gamma$ of the inflaton. In particular, we will show that a power-law dependence on the Higgs background $\Gamma\propto h^n$ does not work. It seems that previous literature on Higgs-modulated reheating have ignored the post-inflationary Higgs decay altogether and thus should be reconsidered. In addition, the quartic self-coupling of the SM Higgs boson generally leads to a rather large local-shape non-Gaussianity  which could already be in conflict with CMB constraints.

We show in this paper that the Higgs-modulated reheating can work if the inflaton decay rate depends on the Higgs background through its dependence on the mass of the decay products. Very often the masses of matter fields are dependent on the Higgs background, so this requirement is rather generic. In particular, we show that a successful CHC scenario can be realized if the inflaton decays dominantly into Higgs-portal scalar particles. In this scenario, the inflaton does not couple directly to any SM particle, and thus the SM signals in CHC receive little ``pollution'' from the free parameters in the inflaton-matter couplings. On the other hand, the potentially large local non-Gaussianity is avoided if the Higgs modulation contributes only a fraction of all density perturbations.

One additional important question is how to tell the CHC scenario from more conventional inflation models. In other words, how can we know that the density perturbation is really from Higgs fluctuations but not the inflaton fluctuation? Unfortunately, this question has no solution in general. But in specific realizations of CHC, such as the Higgs-portal scenario mentioned above, we can make use of the well-understood SM physics to ``calibrate'' the CHC. On the other hand, we should note that, instead of viewing the ``difficult-to-tell'' as a problem, we may also view it as a window to more opportunities. Since after all we haven't discovered a heavy particle through inflation yet. Having CHC as one more discovery channel certainly enhances the chances for such a discovery in the future. 

For the ``calibration'' of the CHC, and also to illustrate the basic techniques of calculating CHC signals, we work out the dominant contribution to the ``clock signals'' in the squeezed bispectrum from $W/Z$ bosons and the top quark. Contrary to the conventional scenario of cosmological collider where large uncertainties could be introduced due to large number of free parameters, the relatively well-understood SM physics allows us to make quite certain predictions about the strength and the shape of the ``clock signals,'' and therefore, we can use these signals to calibrate the CHC. This will be the first step of using CHC as a machine for exploring new physics.

In the rest of the paper, we will review the basic physics of the modulated reheating scenario in Sec.~\ref{sec_overview} and study the evolution of the Higgs field in Sec.~\ref{sec_evol_higgs}. In Sec.~\ref{sec_HMID} we study the Higgs-modulated inflaton decay, and show that it is in general difficult to accommodate Higgs-modulated reheating if the inflaton decays directly into SM particles. On the contrary, the Higgs-modulated reheating has a very simple realization if the inflaton decays dominantly into Higgs-portal scalars, and the SM particles are created only through the thermalization of the Higgs-portal scalars. In Sec.~\ref{sec_CHCsignal} we calculate SM signals in CHC and show that they can be naturally large. Further discussions are in Sec.~\ref{sec_discussions}.

Throughout the paper, we use the $(-,+,+,+)$ signature for the spacetime metric. Both the comoving time $t$ and conformal time $\tau$ will be used. They are related by $a(t)\di t=\di \tau$ where $a(t)$ is the scale factor. The time derivative of a quantity $q$ is denoted by $\dot q\equiv \di q/\di t$ and $q'\equiv\di q/\di\tau$. The Hubble parameter is $H(t)=\dot a(t)/a(t)$. We also use several subscripts to denote quantities at different epochs of cosmic evolution. For instance, $H_\text{inf}$, $H_\text{end}$, and $H_\text{reh}$ denote the Hubble parameter during inflation, at the end of inflation, and at the time of reheating, respectively. The Hubble parameter during inflation is almost a constant and thus we have $H_\text{end}\simeq H_\text{inf}$.

\section{Overview of Modulated Reheating}
\label{sec_overview}

In this section we review the basic idea of the modulated reheating scenario \cite{Dvali:2003em, Kofman:2003nx, Suyama:2007bg, Ichikawa:2008ne}, with a special focus on the case of the SM Higgs boson as the field modulating the inflaton decay. We will describe the mechanism converting the inhomogeneities of the Higgs field into that of the temperature through reheating. The Higgs fluctuation at the time of reheating is assumed known. The evolution of the Higgs field, including its background value and fluctuations, will be considered in the next section. To be specific and also for simplicity, we will assume a standard single-field slow-roll inflation scenario throughout the paper. Generalizations to more complicated scenarios are possible and will be left for future study.

The basic idea of modulated reheating is to generate large-scale density fluctuations by the local perturbation of the reheating temperature. This can be realized if the decay rate of the inflaton is dependent on the background value of some light scalars. Then, the fluctuations of these light scalars will perturb the decay rate, and thus perturb the local expansion history.

Light scalars are ubiquitous in particle physics. In this paper we identify the light scalar modulating the inflaton decay to be the SM Higgs boson, but the general picture reviewed below could also apply to other light scalars having isocurvature perturbations, providing other types of cosmological isocurvature colliders. The Higgs modulated reheating follows the below procedure:

1) During inflation, the Higgs field develops a uniform and slow-roll background, and also a nearly scale-invariant fluctuation on top of the background. Both of them can be generated purely from quantum fluctuation, if the Higgs field is much lighter than the Hubble scale, which is usually the case. 

2) After inflation, the inflaton decays eventually to SM particles and thus heats the universe. The decay rate of the inflaton can be dependent on the Higgs background value. This can be achieved in at least two ways: First, the decay rate can have dependence on the mass of decay products, which in turn depend on the Higgs background. Second, when one couples the inflaton to SM particles in an EFT manner, there are operators involving the Higgs field. In this case the Higgs background will directly control the decay rate.

3) Assuming that the decay rate of the inflaton is much smaller than the Hubble scale $H_\text{end}$ right after inflation, the universe would reheat only at a later time $t_\text{reh}$ when $H(t_\text{reh})\sim \Gamma$. (In general, there are also time dependence in the decay rate $\Gamma$ which we will elaborate below.) Further assuming that the inflaton potential at the bottom is quadratic so that the oscillating inflaton background behaves effectively as cold matter, and that the decay products are radiation, we see that a shift of decay time will alter the rate of local expansion. Therefore, the Higgs fluctuation would be translated to the temperature fluctuation via the Higgs-dependent decay rate.

Now we elaborate the above idea quantitatively. We first split the inflaton $\phi(t,\mb x)=\phi_0(t)+\de\phi(t,\mb x)$ into the space-independent background value $\phi_0$ and the space-dependent fluctuation $\de\phi(t,\mb x)$, and we do the same for the Higgs field $h(t,\mb x)=h_0(t)+\de h(t,\mb x)$. The goal is to express the curvature fluctuation $\zeta(t,\mb x)$ in terms of the Higgs fluctuation $\de h(t,\mb x)$ and also the inflaton fluctuation $\de\phi(t,\mb x)$. This can be done by noting that the local $e$-folding number $N(t_1,t_2;\mb x)$ between time $t_1$ and $t_2$ is defined to be
\begin{equation}
  N(t_1,t_2;\mb x)=\int_{t_1}^{t_2}\di t\,H(t,\mb x).
\end{equation}
Here $H(t,\mb x)\equiv \dot a(t,\mb x)/a(t,\mb x)$ is the local Hubble parameter, and $a(t,\mb x)=\bar a(t)e^{\zeta(t,\mb x)}$ is the local scale factor. Therefore,
\begin{equation}
  N(t_1,t_2;\mb x)=\int_{t_1}^{t_2}\di t\,\ob{H}(t)+\zeta(t_2,\mb x)-\zeta(t_1,\mb x).
\end{equation}
Here $\ob H(t)\equiv \dot{\bar a}(t)/\bar a(t)$, and we identify the integral $\int\di t\,\ob H\equiv \ob{N}$ as the unperturbed $e$-folding number. Then \cite{Starobinsky:1986fxa, Sasaki:1995aw, Lyth:2004gb}
\begin{equation}
  \zeta(t_2,\mb x)=\zeta(t_1,\mb x)+ \de N(t_1,t_2,\mb x), ~~~~~\de N(t_1,t_2,\mb x)\equiv N(t_1,t_2,\mb x)-\ob{N}(t_1,t_2,\mb x).
\end{equation} 

Since we have included the curvature perturbation $\zeta$ in $a(t,\mb x)$, we need to specify how the spacetime is sliced when speaking of the time $t$. Here we choose uniform energy-density slice at all times, and we choose $t_1$ during inflation, and $t_2$ well after reheating has finished. Then, $\zeta(t_1,\mb x)$ will be the curvature perturbation generated during inflation, as in the simplest inflation model, by the inflaton fluctuation. Therefore we have the usual conversion by redefining $t_1$-slice as uniform-curvature slice,
\begin{equation}
  \zeta(t_1,\mb x)=-\FR{H_\text{inf}}{\dot \phi_0}\de\phi(t_1,\mb x),
\end{equation}
where $H_\text{inf}$ is the Hubble scale during inflation and $\dot\phi_0$ is the rolling speed of the inflaton background and is almost a constant during inflation.
The large scale modes already outside the horizon at $t=t_1$ will remain constant $\de\phi\simeq H_\text{inf}/(2\pi)$ in the subsequent evolution until reentering the horizon.

If the expansion histories between $t_1$ and $t_2$ are identical among all local Hubble patches, then $\delta N(t_1,t_2,\mb x)=0$. But, just as we mentioned earlier, Higgs fluctuations plus a Higgs-dependent decay rate $\Gamma(h)$ can perturb the expansion history in different Hubble patch, and therefore introduce nonzero $\de N(t_1,t_2,\mb x)$. To see this more explicitly, we consider a typical case where the inflaton potential is quadratic around the minimum so that the post-inflationary universe is effectively dominated by matter before reheating. Further assuming that the decay product of the inflaton is radiation, then the universe will be radiation dominated after reheating. Let $t_\text{reh}$ be the time when $H=\Gamma$, then we can write,
\begin{equation}
  N(t_1,t_2;\mb x)=\int_{t_1}^{t_\text{reh}}\di t \ob H(t)+\int_{t_\text{reh}}^{t_2}\di t \ob H(t)=\FR{2}{3}\log\FR{t_\text{reh}}{t_1}+\FR{1}{2}\log\FR{t_2}{t_\text{reh}}.
\end{equation}
Here we have used $H(t)=2/(3t)$ for matter domination and $H(t)=1/(2t)$ for radiation domination.
Now we perturb $\Gamma$, and thereby perturb $t_\text{reh}$. From the relation $H=2/(3t)$ $(t\leq t_\text{reh})$ and $H(t_\text{reh})=\Gamma$, we get $\de t_\text{reh}/t_\text{reh}=-\de\Gamma/\Gamma$, and therefore,
\begin{equation}
  \de N=-\FR{1}{6}\FR{\de\Gamma}{\Gamma}.
\end{equation}
Other scenarios with different expansion rates before and after reheating would give different pre-factors, not necessarily $-1/6$. But in general it is some $\order{1}$ number. In this paper we will simply take $-1/6$ for clarity. Combining the inflaton perturbation during inflation and the modulated reheating, we have the curvature fluctuation,
\begin{equation}
\label{zeta}
  \zeta=-\FR{H}{\dot \phi_0}\de\phi -\FR{1}{6}\FR{\de\Gamma}{\Gamma}.
\end{equation}
Correspondingly, the power spectrum of the curvature perturbation is a sum of two parts. (We assume that the mixing between $\de \phi$ and $\de h$ is small.)
\begin{align}
\label{PS}
  &P_\zeta=P_\zeta^{(\phi)}+P_\zeta^{(h)}, 
  &&P_\zeta^{(\phi)}=\FR{H^2}{\dot\phi_0^2}P_{\de\phi},
  &&P_\zeta^{(h)}=C_h^2 P_{\de h},
  &&P_{\de\phi}=P_{\de h}=\Big(\FR{H}{2\pi}\Big)^2.
\end{align}
The factor $C_h$ depends on the nature of reheating and the post-inflationary evolution of the Higgs field, which we shall calculated in Sec.\;\ref{sec_HMID}. Depending on the relative sizes of $P_\zeta^{(h)}$ and $P_\zeta^{(\phi)}$, we can have 3 different scenarios:
\begin{enumerate}
  \item $P_\zeta^{(\phi)}\gg P_\zeta^{(h)}$: The inflaton fluctuations dominates the primordial fluctuations, and we have the usual inflation scenario.
  \item $P_\zeta^{(\phi)}\ll P_\zeta^{(h)}$: The Higgs fluctuations dominate, and we have the Higgs-modulated reheating.
  \item $P_\zeta^{(\phi)}\sim P_\zeta^{(h)}$: Mixed scenario. 
\end{enumerate}
We will show in Sec.~\ref{sec_CHCsignal} that  the Higgs self-interaction will in general lead to large local non-Gaussianity. In the Higgs-fluctuation-dominated scenario $P_\zeta^{(\phi)}\ll P_\zeta^{(h)}$, this local non-Gaussianity is often too large to be consistent with CMB bound. Therefore, we will consider a general mixed scenario where the inflaton and Higgs fluctuations contribute together to the density perturbation. For later use, we define the fraction of Higgs boson in the ``external lines'' of CHC diagrams by the ratio $R_h$,
\begin{equation}
\label{Rh}
  R_h\equiv \bigg(\FR{P_\zeta^{(h)}}{P_\zeta}\bigg)^{1/2}.
\end{equation}

The power spectrum at CMB scale is measured to be $P_\zeta\simeq 2\times 10^{-9}$. In ordinary inflation (Case 1), $P_\zeta$ is generated from the inflaton, and therefore we have the fixed relation between the rolling speed of the inflaton $\dot\phi_0$ and the Hubble scale $H_\text{inf}$, namely $\dot\phi_0=(2\pi P_\zeta^{1/2})^{-1}H_\text{inf}^2\simeq (60H_\text{inf})^2$. In Case 3 of mixed scenario, the inflaton contribution a fraction of $P_\zeta$, namely $P_\zeta^{(\phi)}=(1-R_h^2)P_\zeta$. So, we have 
\begin{equation}
  \dot\phi_0=\FR{H_\text{inf}^2}{2\pi[(1-R_h^2)P_\zeta]^{1/2}}.
\end{equation}

So $\dot\phi_0$ will be larger than it is in ordinary scenario (Case 1). This point is important for the study of CHC, because when we write down effective operators during inflation, the rolling speed $\dot\phi_0^{1/2}$ will be the lower bound of the cutoff scale $\Lambda$, $\Lambda\gtrsim\dot\phi_0^{1/2}$, as required by perturbative unitarity. Therefore, we see that the unitarity bound on the cutoff scaler for CHC is higher than the ordinary cosmological collider.

\section{Evolution of the Higgs Field}
\label{sec_evol_higgs}

In this section we study the evolution of the Higgs field, including its background $h_0$ and the fluctuation $\de h$. The basic picture is that, during inflation (Sec.~\ref{sec_higgs_inf}), fluctuations of the Higgs field can be produced through quantum fluctuation. The longest mode leaving the horizon at the beginning of observable inflation will look like a uniform background $h_0$, while the shorter modes will be like space-dependent fluctuations. After inflation (Sec.~\ref{sec_higgs_postinf}), the Higgs will start oscillating in its quartic potential well. The Hubble friction will reduce the amplitude of this oscillation. It is of great importance to understand how fast this reduction is for the study of Higgs modulated reheating.

Before entering the detailed analysis, we should note that the fluctuations of interest are of CMB or LSS scales, which are much longer than the Hubble radius of the universe between the end of inflation $t_\text{end}$ and reheating $t_\text{reh}$. Therefore, the fluctuation field can be treated effectively as a constant within a Hubble patch during this epoch.

\subsection{Higgs Dynamics During Inflation}
\label{sec_higgs_inf}

We assume a high scale inflation scenario for the study of this paper. By high scale we mean that the Hubble parameter $H_\text{inf}$ during inflation is much higher than the electroweak scale $H_\text{inf}\gg\order{100\text{GeV}}$. The original negative quadratic term in the SM Higgs potential can then be neglected and we are left with a quartic potential with $V(h)=\lam h^4/4$. This corresponds to a classically massless scalar field $h$ with quartic self-coupling. During inflation when the spacetime is de Sitter-like, such a scalar field could develop large quantum fluctuations\footnote{It is possible that a non-minimal coupling between the Ricci scalar and the Higgs term $\xi h^2R$ may break the electroweak symmetry during inflation \cite{Chen:2016hrz,Kumar:2017ecc}. In this case the Higgs mass and the two-point function are modified accordingly. By fine-tuning $\xi$, one can make the Higgs light and be used for modulated reheating. Here we will not consider this possibility in details.} $\la h^2\ra\sim H_\text{inf}^2/\sqrt{\lam}$. This will introduce a dynamical mass $m^2\sim\lam\la\phi^2\ra\sim \sqrt{\lam}H_\text{inf}^2$ to the originally massless scalar $h$. One can think of this mass as a thermal mass in inflation (though the thermal distribution is not the conventional Bose-Einstein because of the redshifts from the expansion of the universe), since the Hawking radiation coming from the dS horizon carries a temperature $T\sim H_\text{inf}^{-1}$. To see this point more quantitatively, we can calculate the 2-point function the scalar field $h$ with loop corrections. One can do this either in the real-time formalism or by doing Wick rotation into the Euclidean dS. The zero-mode approximation in the Euclidean dS allows us to sum over all loops. The result is \cite{Chen:2016hrz}
\begin{equation}
  m_h^2=\sqrt{\FR{6\lam}{\pi^3}}H_\text{inf}^2.
\end{equation}

For small $\lambda$, and considering the extra powers of $\pi$ suppression, the Higgs can be light enough, such that its fluctuation does not depart from scale invariance very much. This contribution, combined with the time dependence of $H$, and the inflaton fluctuation as indicated in the first term in (\ref{zeta}), should be possible to fit the observed tilt of the nearly scale-invariant power spectrum. Moreover, to work out a complete story, it is important to consider the post-inflationary evolution of the Higgs field, which we will do below.

\subsection{Post-Inflationary Evolution of the Higgs Field} 
\label{sec_higgs_postinf}

We assume that the Higgs field is light during inflation, $ m_h\ll H$, which is easily satisfied. Consequently, light Higgs acquires large fluctuations with typical amplitude of $\order{H}$. After inflation, this background will eventually roll back to its origin and also decay to SM particles whenever it can. Therefore, for the purpose of modulated reheating, it is important to check that the Higgs fluctuation at the time of reheating is still large enough. 

Again we assume $H_\text{inf}\gg v\simeq 246$GeV. Then, right after inflation and before the completion of reheating, the universe is cool enough that the Higgs potential is dominated by its quartic term. The Higgs background $h_0$ then follows the equation,
\begin{equation}
\label{heqfull}
  \ddot h_0(t)+(3H+\Gamma_h)\dot h_0(t)+\lam h_0^3(t)=0.
\end{equation}
Here $\Gamma_h$ is the decay width of the Higgs field to SM particles, which is proportional to Higgs background value $h_0$. The ratio $\ga_h\equiv \Gamma_h/h_0$ is almost a constant at different energy scales and thus we can take its value at the electroweak vacuum, where $h_0\simeq 246$GeV and $\Gamma_h\simeq 4$MeV \cite{Tanabashi:2018oca} so that $\gamma_h\simeq 1.6\times 10^{-5}$. Therefore, for Higgs field with background value $h_0\sim\order{H}$, the perturbative decay of the Higgs field can be neglected compared to Hubble friction. RG running of $\ga_h$ only brings $\order{1}$ corrections and thus is unimportant.

We further assume that the inflaton potential at the bottom is nearly quadratic so the universe before the completion of reheating is effectively dominated by cold matter, and therefore we can take $H=2/(3t)$.  Then, the equation of Higgs background simplifies to
\begin{equation}
\label{heq}
  \ddot h_0(t)+\FR{2}{t}\dot h_0(t)+\lam h_0^3(t)=0.
\end{equation}
This equation shows that the Higgs field after inflation will start oscillating in its quartic potential well. The oscillation amplitude will decay due to the Hubble friction. The exact solution to (\ref{heq}) should be found numerically, but the qualitative feature of $h_0$ evolution can be inferred from the following simple argument. 

It is known that an oscillating classical field in a quartic potential behaves effectively like radiation. Therefore we expect that the energy density of the Higgs field $\rh_h=\frac{1}{4}\lam h_0^4$ will decay according to $a^{-4}$. On the other hand, as mentioned above, the energy density right after inflation is dominated by the inflaton potential which we assumed to be quadratic, and therefore we know that the universe is effectively dominated by matter, and therefore $a\propto t^{2/3}$. Consequently, we see that the background value $h_0$ of the Higgs field decays according to $h_0\propto \rho_h^{1/4}\propto a^{-1}\propto t^{-2/3}$.  
 
\begin{figure}[tbph]
\centering
\vcenteredhbox{\includegraphics[width=0.47\textwidth]{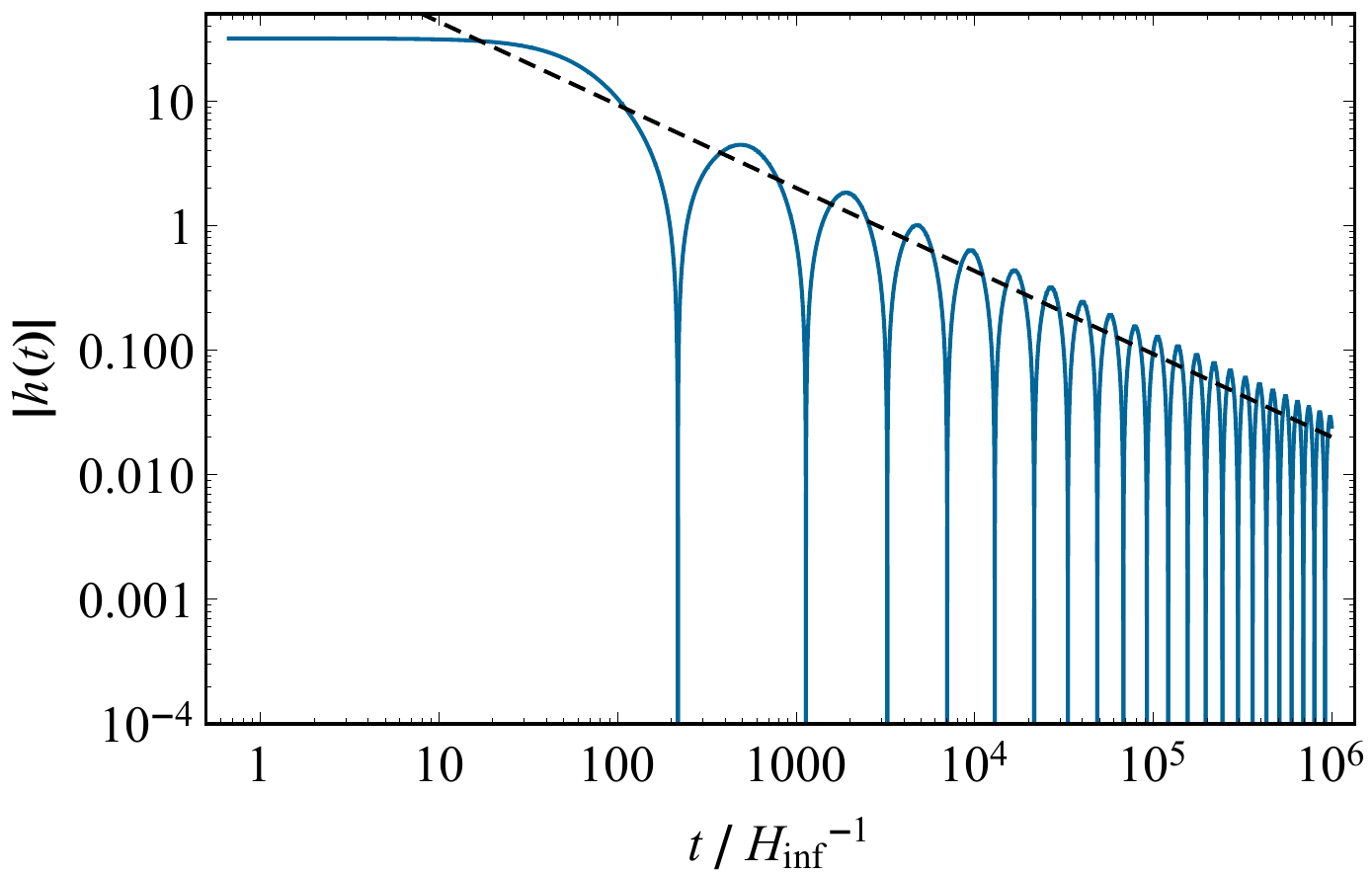}}
\vcenteredhbox{\includegraphics[width=0.47\textwidth]{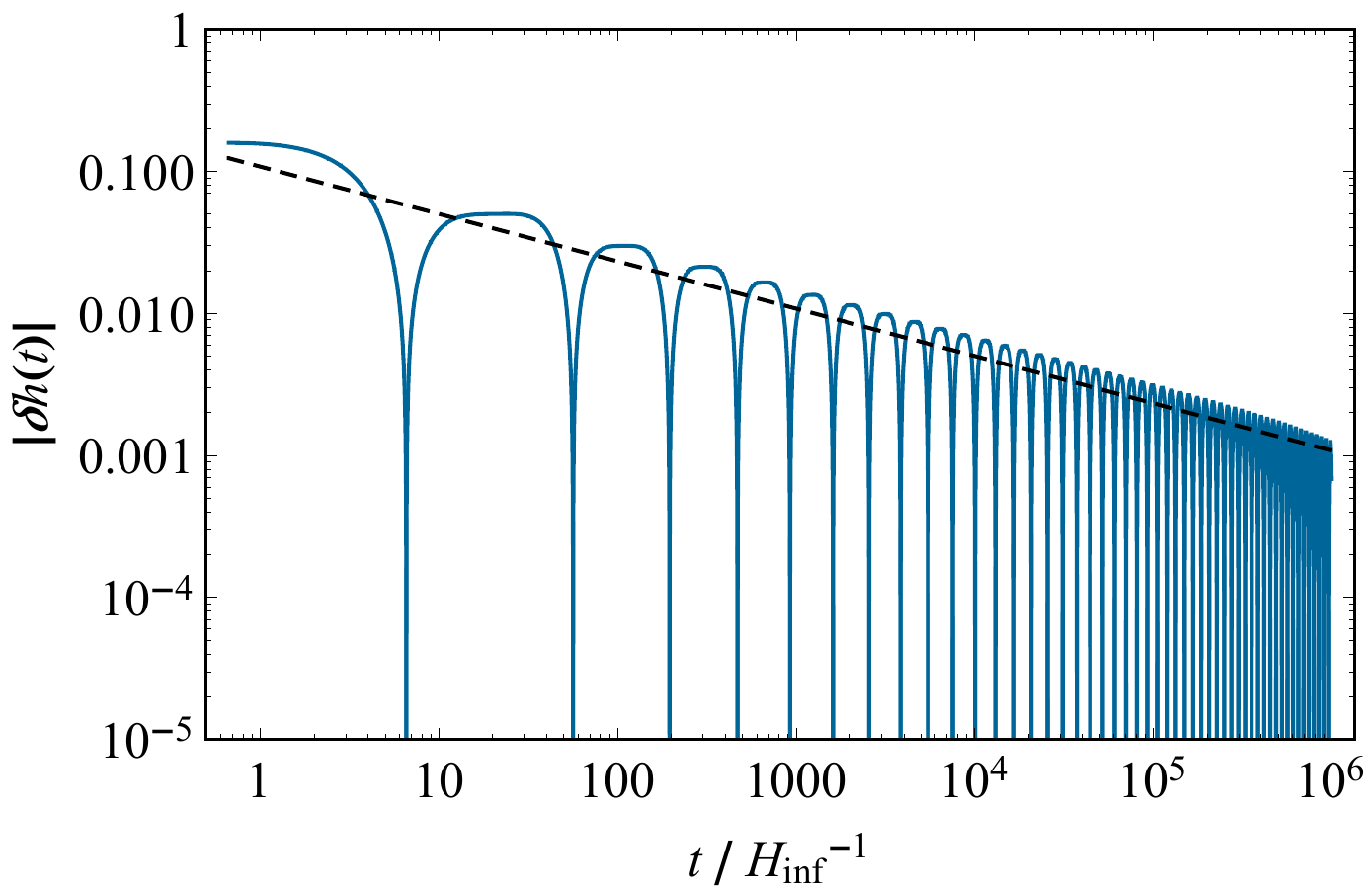}}
\caption{Left: $|h_0(t)|$ as a function of time $t$. Right: $|\de h(t)|$ as a function of time $t$. In both panels the blue solid curves represent numerical solutions of (\ref{heq}), while black dashed curves correspond to analytical fits (\ref{htfit}). For illustration we take $\lam=0.01$ and $\de h_\text{ini}=0.05$.}
\label{Fig_ht}
\end{figure}

With the above picture in mind, we solve the equation (\ref{heq}) numerically. From the numerical solution in Fig.~\ref{Fig_ht} we see that the Higgs background does oscillate with decaying amplitude $h_0\sim t^{-2/3}$. 

In addition, we are also interested in the time evolution of the Higgs fluctuation $\de h(t)$. As noted previously, all Higgs fluctuations we are interested in are outside the horizon between the end of inflation and the completion of reheating, i.e., they look like constant background within a Hubble patch. Therefore, the time evolution of $\de h(t)$ can be found simply by solving (\ref{heq}) twice, one with initial condition $h_1(t_\text{ini})=h_0$ and the other with $h_2(t_\text{ini})=h_0+\de h_\text{ini}$. Then the time evolution of $\de h$ is simply given by $\de h(t)=h_2(t)-h_1(t)$. In all cases we take $\dot h=0$ initially since we know that Higgs field is slowly rolling during inflation, and giving Higgs background a small initial velocity will not significantly alter its subsequent evolution.

We show a numerical solution of $\de h$ also in Fig.~\ref{Fig_ht}. From the numerical solution to (\ref{heq}) it turns out that the amplitude of the background oscillation $h_0$ and of the fluctuation $\de h(t)$ are well fit by
\begin{align}
\label{htfit}
&h_0(t)\sim \lam^{-1/3}H_\text{inf} (H_\text{inf}t)^{-2/3}, 
&&\de h(t)\sim \lam^{-1/6}( H_\text{inf} t)^{-1/3}\de h_\text{ini},
\end{align}
as long as $t$ is not too large. The behavior $h_0\sim t^{-2/3}$ is just as expected, but the behavior $\de h\sim t^{-1/3}$ shows that the fluctuation of Higgs field decays a bit slower than the background. This is from the nonlinear nature of the equation (\ref{heq}) and certainly it cannot hold for arbitrarily long. If we evolve $\de h$ long enough then more complicated behavior will appear. The example in Fig.~\ref{Fig_ht} shows that the $\de h\sim t^{1/3}$ behavior holds well within $10^6$ times of the initial Hubble (time) scale. This is long enough for our following analytical estimate.\footnote{At first sight, the different scaling in background and fluctuation seems puzzling. Because the splitting between background and fluctuation is arbitrary and we would have split them in a different way. But note that the scaling $h_0 \sim t^{-2/3}$ is an approximate solution and the precise solution is not a power law. The subleading terms can be as large as $\delta h$ and modify the scaling. To make this point clear, consider a toy function $f=t+\alpha t^2$, where $\alpha$ is small. We may have split $f_0=t-\alpha t^2$ and $\delta f=2\alpha t^2$. Then the scaling $f_0\sim t$ is approximate with subleading correction scales as $\alpha t^2$, but clearly there is nothing wrong. The amplification of difference in initial condition is a general feature for a broad class of non-linear systems. As we have mentioned in the text, the approximation cannot hold for arbitrarily long before the peculiar term catches up. But for the period of time we are interested in, it's safe to trust this approximation.}

\section{Higgs-Modulated Inflaton Decay}
\label{sec_HMID}

Now we consider the decay of the inflaton after inflation. The inflaton can decay in many different ways. As a simple example, we will show that the inflaton decay into Higgs-portal scalar particles can provide a viable realization of Higgs-modulated reheating. We will also consider possibilities of inflaton decay into SM particles. We will show that it is in general not easy to realize Higgs-modulated reheating if the inflaton only decays into SM particles. 

Some general remarks are in the following, before we proceed with detailed analysis. 

As we mentioned earlier, the Higgs-modulation of the inflaton decay happens at the time when reheating is complete, namely when the decay rate $\Gamma$ is comparable to the Hubble scale. Then, a perturbed decay rate will either advance or delay reheating and thus generates the observed density perturbation. For this purpose, an instantaneous reheating right after inflation (namely the decay rate is very large and comparable to the Hubble scale at the end of inflation) is certainly not favorable, because the effect of perturbed decay rate would be too weak.

Therefore, it is preferable to have a delayed reheating scenario, where the decay rate $\Gamma$ is significantly smaller than the Hubble scale at the end of inflation. In this case, only when the Hubble scale drops (typically according to $H\sim t^{-1}$) down to the scale of $\Gamma$, reheating could then be completed. 

However, we also do not want a too slow reheating for a successful generation of density perturbation through Higgs modulation. This is because the Higgs background field will decrease after inflation, as studied above. If reheating is too slow, then the Higgs background would decrease to too small value to generate correct amount of density perturbation. So a successful Higgs-modulated reheating would put a lower bound on the decay rate $\Gamma$ which could be quite nontrivial in some cases. 

In the literature, the modulated reheating scenario is usually studied with two assumptions being made.  First, It is usually assumed that the decay rate $\Gamma$ is a constant independent of time. Second, the decay rate is dependent on the background value of some light scalar field $\sigma_0$, and in literature it is usually assumed that the dependence follows a power law, $\Gamma(\si_0)\propto \si_0^n$. These two assumptions hold simultaneously only when the background light field does not evolve significantly after inflation, which is possible if $\si_0$ is in a flat direction. But this is certainly not the case if we identify $\si_0$ to be the SM Higgs boson, as we have shown previously that the Higgs background value decrease according to $h_0(t)\sim t^{-2/3}$ after inflation. Therefore, we must take account of the fact that the decay rate $\Gamma$ is also a function of time.

Furthermore, it is also easy to see that a power-law dependence $\Gamma\propto h_0^n$ has difficulty in generating a successful modulated reheating. This is because, to have the power-law dependence $\Gamma\propto h_0^n$, one needs to couple the inflaton $\phi$ to at least one single power of Higgs field in an operator like $\mathcal{O}\sim \phi h\cdots$ with possible spacetime derivatives acting on these fields. Then we see that the inflaton $\phi$ can decay to particles represented by ``$\cdots$'' in $\mathcal{O}$ with a decay rate $\Gamma\propto h_0^2$. That is, the smallest nonzero exponent $n$ in the power-law dependence $\Gamma\propto h_0^n$ will be $n=2$. Then, from $h_0(t)\propto t^{-2/3}$ we see that the decay rate $\Gamma$ will decrease with time no slower than $t^{-4/3}$. This decrease is already faster compared with the Hubble scale $H(t)\sim 1/t$, so it is impossible for $H(t)$ to catch up $\Gamma$ at a later time to achieve the successful reheating. 

Two remarks to complete the above argument. First, we may expect that the inflaton couples to at least a pair of Higgs fields since the Higgs doublet carries gauge charges. But this does not contradict the above assumption of linear $h$-dependence in $\mathcal{O}$. Consider for instance the operator $\mathcal{O}\sim \phi\mb H^\dag \mb H \psi^2$ where $\psi $ is arbitrary matter field, one can find a term $\phi h_0\de h\psi^2$ when evaluating $\mathcal{O}$ on the Higgs background $h_0$ and this term will provide the linear $h$-dependence.  Second, one can also consider a polynomial dependence in $\Gamma(h_0)\sim \Gamma_0+\Gamma_2h_0^2+\cdots$ which can be easily realized in simple particle models. In this scenario, the inflaton decay rate does not drop quickly to zero due to the constant piece $\Gamma_0$, while the desired Higgs modulation is achieved through the $\Gamma_2h_0^2$ term. However, this type of dependence will not generate correct amount of density perturbations, which can be seen after going through the analysis similar to what we will do in the next subsection.

Given the fact that $\Gamma\sim h^n$ with power-law dependence decreases faster than Hubble scale $H$, we could imagine that we can have a decay rate $\Gamma$ initially slightly larger than $H$ at the end of inflation, so we have an instantaneous reheating. The term ``instantaneous'' is actually misleading here because reheating would take at least take a duration of $\order{\Gamma^{-1}}$ to finish. But since $\Gamma$ is decreasing fast, we could imagine that reheating is terminated before it is completed. The rest of inflaton could decay through other much slower channels, e.g., through a gravitational coupling from operators suppressed by Planck scale. In this way, a perturbation in the Higgs field $h_0$ would lead to advanced/delayed termination of reheating, and thus a perturbation in the thermal expansion of the local universe. While more quantitative study is needed to justify whether this scenario is possible or not, here it suffices to mention that a decay rate initially larger than Hubble would typically require a rather large coupling between the inflaton to its decay products, which could be easily constraint from physics during inflation, such as the back reaction to the inflaton potential or the perturbative unitarity. Also, before the further decay of the inflaton, the oscillatory inflaton background will dilute the existing radiation, which suppresses the effect of modulated termination of reheating. Therefore, we will not consider this case any more in this paper.
 
Then it seems difficult for the SM Higgs boson to modulate reheating. Fortunately, the decay rate $\Gamma$ can depend on the Higgs background value $h_0$ through the kinematic factor rather than through a power law. Consider a simple example where the inflaton decays to a pair of massive gauge bosons $A_\mu$ and the gauge boson gets the mass from the Higgs background. Let the decay be from the coupling $\phi F\wt F/\Lambda$ where $F_{\mu\nu}=\pd_\mu A_\nu-\pd_\nu A_\mu$ and $\wt 
F$ being the dual of $F$. The decay rate will have the following form,
\begin{equation}
  \Gamma(\phi\to AA)\propto \FR{m_\phi^3}{\Lambda^2}\bigg(1-\FR{4m_A^2}{m_\phi^2}\bigg)^{3/2}.
\end{equation} 
Then we see that $\Gamma$ depends on the Higgs background $h_0$ where the gauge boson $A$ get its mass, $m_A\propto h_0$. In this way, the Higgs background fluctuation will perturb the decay rate and thus modulate the reheating process, while the decay rate itself does not decrease significantly, so that a delayed reheating is easily achieved: We just need to have a decay rate $\Gamma$ initially smaller than $H$, and wait until $H$ drops to the value of $\Gamma$. 

In the following we will make use of this kinematic dependence on the Higgs background value to modulate reheating. As mentioned at the beginning of this section, there are many ways to make the inflaton decay. We will only consider dim-5 operators respecting the shift symmetry $\phi\to\phi\,+\,$const.\ which frees us from worrying about the back reaction to the inflaton potential. We will begin with the dim-5 couplings to SM particles (fermions, gauge bosons and the Higgs boson) and show why it is quite difficult to modulate reheating through these channels. Then we will move on to the Higgs-portal scalar channels, where the inflaton decays to Higgs-portal scalar bosons, which turns out to be the easiest way for the CHC to work.

\subsection{SM Fermion and Gauge Boson Channels}
\label{sec_fer_gauge_channel}

The only dim-5 couplings between the inflaton and SM gauge bosons and fermions respecting the shift symmetry are the following,
\begin{equation}
\label{dim5coup}
  \Delta\ld =  \sum_i \FR{1}{\Lambda_{f,i}} \bar f_i(\sla\pd \phi)\ga^5 f_i + \sum_\al\FR{1}{\Lambda_{g,I}}\phi F_I \wt F_I,
\end{equation}
where $f_i$ represents all SM fermions, $F_I$ represents the field strengths of SM gauge fields $A_I$, $\Lambda_{f,i}$ and $\Lambda_{g,I}$ are corresponding cut-off scale. These are axion-type couplings, and indeed, axion-like particles can be perfect inflaton candidate. 

The two-body decay rates are
\bgs
\label{decayrates}
\begin{align}
\label{fermionrate}
  &\Gamma(\phi\to f_i\bar f_i)=\FR{1}{2\pi\Lambda_{f,i}^2}m_\phi m_{f_i}^2\bigg(1-\FR{4m_{f_i}^2}{m_\phi^2}\bigg)^{1/2},\\
\label{gaugebosonrate}
  &\Gamma(\phi\to A_I A_I)=\FR{1}{4\pi\Lambda_{g,I}^2}m_\phi^3\bigg(1-\FR{4m_{A_I}^2}{m_\phi^2}\bigg)^{3/2}.
\end{align}
\eds
We note that the decay rates here are for one species in the final states, and it is understood that factors from internal degrees of freedom should be included. For example, an additional factor of 3 should be included in (\ref{fermionrate}) when $f_i$ is a quark, and an additional factor of $8$ should be included in (\ref{gaugebosonrate}) when $A_I$ represents the gluon.

For fermionic decay channels, the two-body decay is dominated  by the heaviest $f_i$ satisfying $m_{f_i}<m_\phi/2$. But there can be cascade decay of heavy fermions, too. These decays will all be controlled by the Higgs background value.

A first observation here is that the fermionic decay rates depend on the Higgs background $h_0$ through a power law, $\Gamma\sim m_f^2\sim h_0^2$. The discussion at the beginning of this section shows that this channel is of no use for modulated reheating if we want a delayed reheating scenario, because the decay rate decreases faster than the Hubble scale. Therefore we will no longer consider this channel any more in this paper. But before moving on to the gauge boson channel, let us also mention that the fermion channel cannot be used in the ``terminated reheating'' scenario mentioned above either, which requires a decay rate initially larger than the Hubble scale. To see it, we note that the Higgs background $h_0\sim H_\text{inf}$ initially, and let us consider the top quark without further suppression from Yukawa coupling. Then, the decay rate is, 
\begin{equation}
\Gamma(\phi\to f\ob f)\sim\FR{ m_\phi H_\text{inf}^2}{\Lambda_f^2}.
\end{equation}
The cutoff scale $\Lambda_f$ cannot be smaller than the unitarity bound, $\Lambda_f\gtrsim\dot\phi_0^{1/2}$ which is typically much higher than the Hubble scale (cf.\ discussion at the end of Sec.~\ref{sec_overview}). Assuming that inflation and reheating are described by the same effective field theory (EFT), we see that $\Gamma(\phi\to f\ob f)$ can never be larger than the Hubble scale. 

Now we turn to the gauge boson channel. Apart from the kinematic factor, the decay rate is approximately  
\begin{align}  
  \Gamma(\phi\to AA)\sim \FR{m_\phi^3}{\Lambda_g^2}.
\end{align}
With the constraint $\Lambda_g > m_\phi$ and $\Lambda_g > \dot\phi_0^{1/2}$ in mind, we see that the  decay rate of the gauge boson channel can be either greater or smaller than the Hubble scale at the end of inflation. Of course, to achieve a delayed reheating, we prefer a smaller $\Gamma$ and thus a higher cutoff scale.  

In the gauge boson channel, the fluctuations in $h$ will generate fluctuations in $\de \Gamma$ through the dependence in the kinematic factor,
\begin{align}
\label{dGammaGaugeboson}
  \de\Gamma\simeq -\sum_I\FR{3 g_I^2}{4\pi}\FR{m_\phi\bar h}{\Lambda_{g,I}^2}\de h,
\end{align}
where we have assumed $m_{A_I}=\frac{1}{2}g h_0\ll m_\phi$. To get an idea of how large the curvature fluctuation can be generated in this channel, let's further assume that $\Lambda_{g,I}$ and $g_I$ are the same for all species, then,
\begin{equation}
\label{zetah}
  \zeta =-\FR{1}{6}\FR{\de\Gamma}{\Gamma}\bigg|_{t=t_\text{reh}}=\FR{g^2h_0(t_\text{reh})\de h(t_\text{reh})}{2m_\phi^2},
\end{equation}
To see how large curvature fluctuations can be generated from the Higgs-modulated decay,  we compute $\la \zeta^2\ra$ as follows.
\begin{align}
  \la \zeta_k^2\ra\simeq \bigg[\FR{g^2h_0(t_\text{reh})}{2m_\phi^2}\bigg]^2\big\la \de h_k^2(t_\text{reh})\big\ra\simeq \bigg[\FR{g^2}{2m_\phi^2}\FR{1}{\lam^{1/2}t_\text{reh}}\bigg]^2\FR{H_\text{inf}^2}{2k^3}.
\end{align}
From the relation $\la \zeta^2\ra=(2\pi^2/k^3)P_\zeta $ and $P_\zeta \simeq 2\times 10^{-9}$, we can determine the reheating time $t_\text{reh}$. Expressing the result in terms of the decay width $\Gamma=2/(3t_\text{reh})$, we have,
\begin{equation}
\label{reqGammaGB}
  \Gamma=\FR{8\pi\lam^{1/2}}{3g^2}\Big(\FR{m_\phi}{H_\text{inf}}\Big)^{2}\sqrt{P_\zeta }H_\text{inf}.
\end{equation}
This can be translated to an expression for $\Lambda_g$ (treating all $\Lambda_{g,I}$ equal to $\Lambda_g$), 
\begin{equation}
\label{Lambdag}
  \Lambda_g=\sqrt{\FR{3Ng^2}{32\pi^2\lam^{1/2}}}P_\zeta^{-1/4}\sqrt{m_\phi H_\text{inf}}.
\end{equation}
To get an idea of how large the cutoff is needed to be, we take an example of $N=4$, $g=0.6$ (taken as the $SU(2)_L$ gauge coupling, and thus we are ignoring decays into photons and gluons, since they are massless and not helpful for kinematic modulation of reheating), and $\lam=0.01$ (the Higgs self-coupling at the inflation scale). Then we have,
\begin{equation}
\label{LambdaValue}
  \Lambda_g\simeq 175H_\text{inf}\bigg(\FR{m_\phi}{10H_\text{inf}}\bigg)^{1/2}.
\end{equation}
Therefore, with $m_\phi$ larger than the Hubble scale during inflation, we will have a cutoff quite close to the unitarity bound $\Lambda_g\gtrsim \dot\phi^{1/2}_0$. (Recall that $\dot\phi_0^{1/2}\gtrsim 60(1-R_h^2)^{-1/2}H_\text{inf}$ when Higgs fluctuations contribute $R_h$ of the density perturbation in terms of the amplitude.) For models with $m_\phi\gg H_\text{inf}$, we can have much higher cutoff scale $\Lambda$. 

Some $\order{1}$ corrections to the above estimate could arise if we consider more realistic situation. For example, in SM, only $W$ and $Z$ are massive among all gauge bosons. Therefore only $\phi\to WW$ and $ZZ$ channels contribute nonzero $\de\Gamma$ in (\ref{dGammaGaugeboson}). On the other hand, all gauge bosons, including 8 colored gluons and the photon, would contribute to $\Gamma$. Taking this into account and rederiving (\ref{zetah}), it's easy to see that the net effect is to replace $m_\phi$ in every formula from (\ref{zetah}) to (\ref{reqGammaGB}) by an effective $\wh m_\phi$, given by,
\begin{equation}
  \wh m_\phi^2=\bigg(\sum_{I=\text{all}}\Lambda_{g,I}^{-2}\bigg/\sum_{I=Z,W^\pm}\Lambda_{g,I}^{-2}\bigg)m_\phi^2.
\end{equation}
If we again take all $\Lambda_{g,I}$'s equal, then we have $\wh m_\phi^2=4m_\phi^2$. We should also replace $m_\phi$ in (\ref{Lambdag}) by $m_\phi/4$, and take $N=12$. As a result, the numerical factor in front of (\ref{LambdaValue}) should include another factor of $\sqrt{3}/2$ and should be 151 instead of 175.

As we will show in the next section, while the cutoff  scale (\ref{LambdaValue}) required for generating appropriate amount of density perturbation is barely consistent with the unitarity bound, such a value of cutoff could lead to gauge boson overproduction during inflation \cite{Meerburg:2012id} through the operator (\ref{dim5coup}). Not-too-much gauge boson production will require that the chemical potential $\mu\sim\max\{H_\text{inf},m_A\}$ where $H_\text{inf}$ and $m_A$ are the Hubble scale and the gauge boson mass during inflation, respectively, and the chemical potential $\mu\equiv\dot\phi_0/\Lambda_g$. The overproduction can be seen from the exponential enhancement in the mode function of the gauge field in the presence of a chemical potential. See the appendix for details. Given $\dot\phi_0> (60H_\text{inf})^2$, we see that it is desirable to have $\Lambda_g$ at least $\order{10^2}$ larger than the value in (\ref{LambdaValue}). 

There are ways to make $\Lambda_g$ larger. One obvious way to raise the cutoff is to have a larger inflaton mass during reheating as can be seen from (\ref{LambdaValue}). Too large inflaton mass at the bottom of the potential well may bring difficulty in the model building of single field inflation. Also, one cannot make the inflaton mass $m_\phi$ higher than the cutoff $\Lambda_g$. Multi-field inflation scenarios such as hybrid inflation \cite{Linde:1993cn} may easily achieve a large $m_\phi$ (in which context $m_\phi$ is actually the mass of the waterfall field). Also, if different stages of inflation are described by different validity regimes of EFT, the constraints on $\Lambda_g$ may get relaxed. To keep the model simple, we will not explore these multi-field or multi-EFT possibilities here.

One can also consider other decay channels in new physics scenario to speed up the decay and thus raise the cutoff further. However, we must keep in mind that presence of additional decay channels could weaken the dependence on the Higgs background by giving a larger denominator to (\ref{zetah}), unless these new channels themselves are dependence on Higgs background. We leave detailed studies on new-physics channels to future works. 

The general conclusion of this subsection is that it is not possible to use SM fermion channel to modulate reheating. It is possible to make use of the gauge boson channel. However, to generate the right amount of primordial perturbation would require a rather low cutoff $\Lambda_g$ that is easily inconsistent with the lower bound from the gauge boson production during inflation. Therefore the viable parameter space for the gauge boson channel is also quite restricted.

\subsection{Higgs Boson Channel}
\label{sec_higgsChannel}

Given the difficulty of achieving modulated reheating in the SM gauge boson and fermion channels, now we move on to the SM Higgs channel. At dim-5 level, we can write down two independent real couplings between the inflaton and the Higgs boson respecting the shift symmetry,
\begin{align}
\label{dim5Higgs}
  \Delta\ld=&~\FR{2}{\Lambda_{hr}}(\pd_\mu\phi)\text{Re}\big(\mb H^\dag \D^\mu \mb H\big)+\FR{2}{\Lambda_{hi}}(\pd_\mu\phi)\text{Im}\big(\mb H^\dag \D^\mu \mb H\big)\n\\
  =&~\FR{1}{\Lambda_{hr}}(\pd_\mu\phi)h\pd^\mu h+  \FR{g}{2c_W\Lambda_{hi}}(\pd_\mu\phi)Z^\mu h^2.
\end{align}
Here $\Lambda_{hr}$ and $\Lambda_{hi}$ are cutoff scales of the two couplings, respectively, and $c_W=\cos\theta_W$ with $\theta_W$ being the Weinberg angle.
In the second line we have taken the unitary gauge $\mb H=(0,\frac{1}{\sqrt 2}h)^T$ where all would-be Goldstone bosons are gauged away.  

For the analysis of inflaton decay, we will only consider the $\Lambda_{hr}$-term. The contribution from the $\Lambda_{hi}$-term is similar, and we will omit it for simplicity. However, for the ``cosmological collider'' signals to be considered in the next section, the coupling $\Lambda_{hi}$ will be very interesting since it will introduce $Z$-exchanging tree-level diagrams.

The $\Lambda_{hr}$-term in (\ref{dim5Higgs}) modifies the dynamical evolution of the Higgs field, both during inflation and after. For now we will proceed as if the previous results on the Higgs evolution are still applicable and will justify this treatment later.

The two-body decay rate of $\phi\to hh$ from the $\Lambda_{hr}$-term is
\begin{align}
  \Gamma(\phi\to hh)=\FR{1}{16\pi\Lambda_{hr}^2}m_\phi^3\bigg(1-\FR{4m_h^2}{m_\phi^2}\bigg)^{1/2}.
\end{align}
The Higgs-dependence in the decay width from this channel is very similar to that in the gauge boson channel, since the Higgs mass during this period is
\begin{equation}
m_h^2=6\lam h_0^2.
\end{equation} 
So we can repeat our analysis for the gauge boson channel in the last subsection. We would expect that the resulting cutoff scale $\Lambda_{hr}$ would be rather close to the unitarity bound. One important difference, though, is that both operators in (\ref{dim5Higgs}) do not lead to copious particle production during inflation, and thus the cutoff scales $\Lambda_{hr}$ and $\Lambda_{hi}$ are free from a much stronger bound constraining the gauge boson channel.

The rest of the analysis is in parallel with the one in the last subsection. First, we write down the curvature perturbation generated at the time of reheating,
\begin{equation}
  \zeta=-\FR{1}{6}\FR{\de \Gamma}{\Gamma}\simeq\FR{4\lam h_0\de h}{m_\phi^2}\bigg|_{t=t_\text{reh}},
\end{equation}
where we assumed that $m_\phi\gg m_h$. From this we can express the decay rate in terms of the power spectrum $P_\zeta$ as
\begin{equation}
  \Gamma=\FR{\pi}{3\sqrt\lam}\FR{m_\phi^2}{H_\text{inf}}P_\zeta^{1/2}.
\end{equation}
So the required cutoff scale $\Lambda_{hr}$ is
\begin{equation}
  \Lambda_{hr}=\FR{\sqrt{3\lam^{1/2}}}{4\pi}P_\zeta^{-1/4}\sqrt{m_\phi H_\text{inf}}.
\end{equation}
Taking $\lam\sim0.01$ and $m_\phi\simeq 10H_\text{inf}$, we have $\Lambda_{hr}\sim21H_\text{inf}$, which is already lower than the unitarity bound. Here it is even impossible to raise the cutoff by giving larger mass to $m_\phi$ because we do not expect $m_\phi$ to be larger than the cutoff scale anyway in single field inflation described by a single EFT. Inclusion of the $\Lambda_{hi}$-term will not change the result significantly. Therefore, it is not likely to realize Higgs-modulated reheating through this channel. 

Although we have reached a no-go result, we will still comment on the effect of the operators  in (\ref{dim5Higgs}) on the Higgs evolution, only to make sure that the above analysis is valid.

It is clear that the $\Lambda_{hi}$-term is irrelevant during inflation. On the other hand, the $\Lambda_{hr}$-term contribute a friction term $(\dot\phi_0/\Lambda_{hr}) \dot h_0$ to $h_0$'s equation of motion. The coefficient $\dot\phi_0/\Lambda_{hr}$ is much greater than the Hubble scale. This makes the Higgs field roll slower than in the usual case. But the Higgs field does not dominate the energy budget during inflation, and thus this slow-down has no consequence on the overall amplitude of the Higgs fluctuation, unlike the case of the ordinary inflation. However, decreasing $\dot h_0$ would bring corrections to the slow-roll parameters and thus corrections to the scale dependence of the power spectrum. As before, we assume that this change can always be compensated by modifying the inflaton potential, so that the overall scale dependence agrees with the CMB measurements. 

After inflation, the term $(\dot\phi_0/\Lambda_{hr})\dot h_0$ is still present in the equation of motion. But it is no longer a friction term since $\phi_0$ is fast oscillating. We assume that the inflaton mass at the bottom of its potential well is much greater than the Hubble scale at the end of inflation. Then, the above term provides a fast oscillating external force to $h_0$. Just like in ordinary forced oscillation, a fast oscillating external force with frequency $\sim m_\phi$ has essentially no effect on the oscillator, whose intrinsic frequency is $m_h\sim \sqrt\lam h_0\ll m_\phi$. Therefore, we can safely neglect this term even it is superficially much greater than the Hubble friction term. One can also understand this by simply taking the average of this over a time scale longer than $m_\phi^{-1}$. Then this oscillating term will be averaged to zero.

\subsection{A Higgs-Portal Scalar Channel}

As we see from the above analysis, the main difficulty of realizing Higgs-modulated reheating for SM-only channels is from the fact that the Higgs background decreases quite fast after inflation, and therefore we need a quick decay of the inflaton, which in turn requires a rather low cutoff scale that is easily inconsistent with various constraints. Now we move on to a simple BSM channel, where the inflaton decays to $N$ heavy real scalars $S_i$ that couple to SM through a Higgs portal. Since the Higgs-portal coupling is not constrained, we can assume a large enough coupling to increase the effect of $\de h$, and we can also make use of the number of scalars $N>1$  to enhance the decay rate. Interestingly, these Higgs-portal scalars can also be dark matter candidate.

For simplicity and also for easy generalization to charged scalars, we add a $\mathbb{Z}_2$ parity to the scalar fields $S_i\to -S_i$. Then the relevant Lagrangian is
\begin{align}
  \Delta\ld=-\FR{1}{2}(\pd_\mu S_i)^2-\FR{1}{2}m_{S0}^2 S_i^2-\al S_i^2|\mb H|^2+\FR{1}{\Lambda_S}(\pd_\mu\phi)S_i\pd^\mu S_i.
\end{align}
Here $\al$-term is the usual Higgs-portal coupling and the last term with dim-5 operator is the coupling to the inflaton through which we allow the inflaton to decay. We assume that $m_{S0}$ is not too smaller than the inflation scale $H_\text{inf}$ so that the background value of $S_i$ sits at the minimum $S_{i0}=0$ during inflation, and thus does not modify the inflaton or Higgs evolution. However, we note that this assumption on $m_{S0}$ is only meant to simplify the analysis, and a scenario with $m_{S0}\ll H_\text{inf}$ can well be viable since $S_i$ can also receive mass from other background contribution during inflation. Interestingly, for $\al\sim\order{1}$, putting $m_{S0}\sim\order{10}\text{TeV}$ will make $S_i$ consistent with a thermally produced dark matter candidate. 

After inflation, the inflaton will gradually decay into $S_i$. At this stage, the mass of the $S_i$ will be dependent on the Higgs background value $h_0$ as,
\begin{equation}
   m_S^2(h_0)= m_{S0}^2+\al h_0^2.
\end{equation} 
At the same time, the Higgs background $h_0$ will also decay. We assume that $m_{S0}^2>\lam h_0^2$ so that the Higgs does not decay into $S_i$. Then, as discussed above, the Higgs background $h_0$ decays mostly due to the Hubble friction, and the result (\ref{htfit}) will apply in this case. Consequently, we can perform an analysis similar to the Higgs channel. The curvature perturbation generated from the Higgs-modulated decay $\phi\to S_iS_i$ is,
\begin{align}
\label{zetaHPS}
  \zeta=-\FR{1}{6}\FR{\de\Gamma}{\Gamma}\simeq \FR{2\al h_0\de h}{3m_\phi^2}\bigg|_{t=t_\text{reh}},
\end{align}
where we again assume $m_\phi>m_S$ to simplify the expression. Then we see that to generate the correct amount of primordial fluctuations requires the cutoff scale $\Lambda_S$ to be 
\begin{equation}
 \Lambda_S\simeq \sqrt{\FR{N\al}{32\pi^2\lam^{1/2}}}P_\zeta^{-1/4}\sqrt{m_\phi H_\text{inf}}.
\end{equation}
Assuming the dimensionless coupling $\al =1$, $N=10$, $m_\phi=10H_\text{inf}$, and again $\lam\simeq 0.01$,  we see that the cutoff scale will be $\Lambda_S\geq 266H_\text{inf}$ and this can be well above the unitairty bound. We can further raise the cutoff in this case by giving a larger mass to $m_\phi$. Again, the dim-5 operator $(\pd_\mu\phi)S\pd^\mu S$ does not lead to copious production of $S$ particles even when the cutoff scale $\Lambda_S$ is as low as the unitarity bound, so we see that the Higgs-portal channel is a viable way to realize the Higgs-modulated reheating.  

After the completion of reheating, the thermal collisions of $S_i$ particles will further produce the Higgs boson and other SM particles. The good thing about this scenario is that the inflaton does not couple directly to SM fields, and thus the SM mass spectrum during inflation only receives corrections from the thermal background, which is calculable and quite definite, and this allows us to find rather predictable and clean signals in the squeezed bispectrum. 
 
\subsection{More Possibilities}

Several assumptions have been made in the study of the inflaton decay in this section, including 1) single-field slow-roll inflation, 2) perturbative inflaton decay, 3) EFT couplings of lowest order between the inflaton and matter fields, respecting the shift symmetry of the inflaton. More possibilities are available without making these assumptions. Below we will mention several of them as interesting alternatives, and leave detailed analysis to future studies.

As we see from the previous discussions, one major difficulty in realizing Higgs-modulated inflaton decay is the too-fast decreasing of the Higgs background, which requires a large decay rate of the inflaton. The large decay rate is usually in tension with various bounds on the cutoff scale $\Lambda$ during inflation. This tension can be released in several ways if we go beyond the three assumptions made above. 

First, if we go beyond the single-field slow-roll paradigm and consider a two-field inflation model such as hybrid inflation \cite{Linde:1993cn}, then the slow-rolling field $\phi_1$ will be different from the field $\phi_2$ that reheats the universe at the end of inflation. In this case there will be no direct relation between the $\phi_1$-SM couplings and $\phi_2$-SM couplings. This will provide us more freedom in realizing the Higgs modulated reheating. Also, even in the framework of single field inflation, during inflation the inflaton may have rolled a longer distance than one EFT can describe, such that inflation and reheating might be described by different EFTs with different cutoffs.

Second, we can imagine that the inflaton decays nonperturbatively through preheating \cite{Shtanov:1994ce, Kofman:1994rk} or tachyonic instabilities \cite{Dvali:1998pa}. This will greatly enhance the decay rate of the inflaton even when the perturbative decay is forbidden. 

Third, we can go beyond the EFT couplings and consider specific inflaton-matter couplings valid at large $\phi$. For instance,  if the inflaton-matter coupling has the following form,
\begin{equation}
  \FR{e^{-\phi/\Lambda'}}{\Lambda}(\pd_\mu\phi)\mathcal{O}^\mu,
\end{equation}
where $\mathcal{O}^\mu$ is some vector operator formed by matter fields, and $\Lambda$, $\Lambda'$ are two independent cutoff scales. This operator can lead to quick inflaton decay when $\phi$ oscillates around $\phi=0$ provided a small cutoff $\Lambda$. On the other hand, this operator will be suppressed during inflation by the large $\phi/\Lambda'$ so a low cutoff $\Lambda$ is free from the unitarity constraint during inflation.

\section{Particle Signals at the Cosmological Higgs Collider}
\label{sec_CHCsignal}

In the previous section we analyzed possible realizations of Higgs-modulated reheating, and we see that the easiest way to modulate the inflaton decay through the SM Higgs field is to introduce a couple of Higgs-portal scalar fields $S_i$. We assume that the inflaton couples only to $S_i$ but not directly to SM particles.  In this section, we are going to explore the CHC signals mainly in this scenario. At the same time, we will also mention possible new signals when the inflaton is allowed to couple directly to SM.

As in the ordinary case of the cosmological collider, by CHC signal we mean the bispectrum of the scalar perturbation in the squeezed limit, namely the three-point correlation of the curvature fluctuation $\la\zeta(\mb k_1)\zeta(\mb k_2)\zeta(\mb k_3)\ra$ where one external momentum, say $k_3$, is much smaller than the other two $k_{1,2}$. The CHC then means the oscillations in the bispectrum as a function of  the ``squeezeness'' parameter $k_1/k_3$. This oscillation is a result of the intermediate massive field attached to the soft external leg and going on-shell. And as usual, by measuring the frequency of the oscillation, we will be able to read the mass of the intermediate massive particle in the unit of Hubble scale $H_\text{inf}$. The special feature of the CHC is that the external legs can be viewed as long-lived Higgs mode, and thus the intermediate massive particles couples to the external lines through Higgs coupling.

The advantage of the Higgs-portal scenario for CHC is clear: the SM fields do not couple directly to the inflaton background, and thus the SM particles' mass will receive contribution only from the Higgs background and the thermal loops, which are predictable. There will be further corrections from the inflaton background if we integrate out the $S_i$ fields and introduce effective couplings between the inflaton and the SM fields. But the contribution from these effective operators is expected to be small. In addition, the couplings between the SM fields and the primordial scalar fluctuations are nothing but the Higgs couplings which are known. Therefore, we expect quite definite predictions of SM signals at CHC. This will be very helpful to tell this scenario of CHC from the ordinary inflation.

In the following we will study the CHC signals of SM particles. We will first consider the simplest Higgs-portal scenario, where inflaton couples only to the Higgs-portal scalars $S_i$. Then we will consider an example of BSM couplings between the inflaton and the SM fields. Throughout the section we will use $H$ to denote the Hubble scale \emph{during} inflation, namely, $H=H_\text{inf}$.

Some general discussions before entering the detail. First, from (\ref{PS}) and (\ref{Rh}), the 3-point function from the Higgs external legs can be written as
\begin{align}
  \la\zeta(\mb k_1)\zeta(\mb k_2)\zeta(\mb k_3)\ra_h'= \bigg(\FR{2\pi}{H} P_\zeta^{1/2}R_h\bigg)^3\la\de h(\mb k_1)\de h(\mb k_2)\de h(\mb k_3)\ra'.
\end{align}
Here and everywhere a prime on the correlator $\la\cdots\ra'$ means that the $\de$-function of momentum conservation is removed. The subscript on the left hand side $\la\cdots\ra_h$ means the contribution from the Higgs.  On the other hand, the shape function $\mathcal{S}(k_1,k_2,k_3)$ of the bispectrum is conventionally defined through 
\begin{equation}
\la\zeta(\mb k_1)\zeta(\mb k_2)\zeta(\mb k_3)\ra'=(2\pi)^4\mathcal{S}(k_1,k_2,k_3)\FR{1}{(k_1k_2k_3)^2}P_\zeta^2.
\end{equation}
Comparing the above two equations, we have
\begin{equation}
\label{shape}
  \mathcal{S}(k_1,k_2,k_3)=\FR{1}{2\pi H^3}R_h^3P_\zeta^{-1/2}(k_1k_2k_3)^2\la\de h(\mb k_1)\de h(\mb k_2)\de h(\mb k_3)\ra'.
\end{equation}
The correlator $\la\de h^3\ra$ can be calculated using the Schwinger-Keldysh formalism, which is very similar to calculating Feynman diagrams \cite{Chen:2017ryl}. The overall size of the shape function $\mathcal{S}$ can be represented by a dimensionless number which is conventionally called $f_\text{NL}$. We also note that there are $\order{1}$ differences in different definitions of $f_\text{NL}$. Very often it is useful just to estimate $f_\text{NL}$ without doing detailed calculation. The estimate goes as 
\begin{align}
\label{Sestimate}
  f_\text{NL}\sim \FR{R_h^3P_\zeta^{-1/2}}{2\pi}\times\text{loop factors}\times\text{vertices} \times\text{propagators}.
\end{align}

Every dimensionful parameter in (\ref{Sestimate}) is measured in the unit of Hubble $H$. The loop factor is the usual one from the momentum integral. For example, we can estimate the 1-loop factor as $1/(16\pi^2)$. The vertices refer to the coupling coefficients from usual Feynman rules, and the propagators can be estimated as $\order{1}$ for fields with mass close to or smaller than Hubble, $m\lesssim H$. For heavier mass, each propagator contributing the non-analytical oscillations will be suppressed by a Boltzmann factor $e^{-\pi (m-\mu)/H}$ with $\mu$ being the corresponding chemical potential. If there is additional source of particle production for the internal line other than the vacuum fluctuation, one should also include corresponding enhancing factors.

Here we use the formula (\ref{Sestimate}) to estimate the local non-Gaussianity. This does not belong to the CHC we are interested in, but it is still important to check that the local non-Gaussianity produced in this scenario is consistent with current constraints from CMB.  

Generally, we expect two sources contributing to local non-Gaussianity. One is from the super-horizon evolution of the Higgs fluctuation due to its self-interaction. The induced cubic interaction with coupling $\sim \lam h_0$ will introduce a secular growth of local non-Gaussianity. For small $\lam$, we expect that the resulting $f_
\text{NL}$ will be proportional to the number of $e$-folds. The second contribution  from the non-linear modulation rate, which means that we have additional non-linear terms $\zeta =z_1 \de h+\frac{1}{2}z_2\de h^2+\cdots$ when converting $\de h$ to $\zeta$. In the Higgs-portal scalar channel (\ref{zetaHPS}) we have $z_2\simeq 2\al N/(3m_\phi^2)$. Combining these two contributions, we can estimate the local non-Gaussianity as, \begin{align}
\label{fnlLocal}
  f_\text{NL}(\text{local})\sim-\order{1}\FR{R_h^3}{2\pi P_\zeta^{1/2}}\lam N_e + \order{1}\FR{R_h^3}{(2\pi)^6P_\zeta}\FR{2\al N}{(m_\phi/H_\text{inf})^2},
\end{align}
We see that both terms in (\ref{fnlLocal}) can be quite large due to the inverse powers of $P_\zeta$. The second term can be suppressed if we have a larger inflaton mass at the end of inflation. The first term gives a more stringent constraint on the parameter space, especially on $R_h$. Given the current CMB constraints $|f_\text{NL}(\text{local})|\lesssim 5$, we have,
\begin{align}
  R_h\lesssim \text{min}\bigg\{0.14\bigg(\FR{\lam}{0.01}\bigg)^{-1/3}\bigg(\FR{N_e}{50}\bigg)^{-1/3},1\bigg\}.
\end{align}
This means that the Higgs fluctuation cannot be the only source of the primordial density fluctuation unless with a very tiny self-coupling $\lam\sim\order{10^{-5}}$. However, we will see below that the CHC signals can still be observably large given the $R_h$ suppression derived here.

\subsection{Signals from SM Couplings}

In this subsection we consider SM signals in the simplest and cleanest scenario, namely the inflaton decay through the Higgs-portal channel. 

In this scenario all particles appear at 1-loop level. From the estimate (\ref{Sestimate}) we see that the largest signal should come from particles with strongest couplings with the Higgs, namely $W/Z$ boson and the top quark. We may also see the Higgs-portal scalars $S_i$ in this way because the scenario under consideration prefers a large coupling between $S_i$ and $H$. However, the mass of $S_i$ is a free parameter, and a mass slightly larger than Hubble by several times could easily render the signal invisible. We actually prefer a large mass for $S_i$ because heavier $S_i$ can help to suppress the coupling between the inflaton and SM fields. Therefore, we will not consider $S_i$ signals further.

\paragraph{Gauge boson signals.} The signals in the squeezed bispectrum from massive SM gauge bosons have been calculated in \cite{Chen:2016hrz}. The corresponding signals at CHC is almost the same as the ``inflaton collider'' signals in \cite{Chen:2016hrz}. Here we will outline the main steps of the calculation and refer readers to \cite{Chen:2016hrz} for more details. We use the diagrammatic representation for each process following Schwinger-Keldysh formalism. See \cite{Chen:2017ryl} for a review.

The Higgs-gauge couplings are from the following Lagrangian,
\begin{align}
\label{DH2}
  \FR{1}{2}|\D_\mu\mb H|^2
\supset \FR{1}{4}g^2(h_0+\de h)^2\Big(W_\mu^+W_{}^{-\mu}+\FR{1}{2c_W^2}Z_\mu Z^\mu\Big) 
\end{align}
Apart from the tree-level gauge boson mass from the $h_0$ background, the $W/Z$ boson also receive masses from infrared-enhanced Higgs loop \cite{Chen:2016hrz}. The loop contribution dominates when Higgs is light, which is true in our case. Therefore we will directly quote the loop mass calculated in \cite{Chen:2016hrz} as following,
\begin{align}
  &m_W^2=\FR{3g^2H^4}{8\pi^2m_H^2},
  &&m_Z^2=m_W^2/c_W^2,
\end{align}
where $c_W=\cos\theta_W$ and $\theta_W$ is the Weinberg angle. From the couplings in (\ref{DH2}) we see that the following two 1-loop diagrams contribute to the 3-point function of the Higgs fluctuation.
\begin{equation}
\label{fd_vectorloop}
  \parbox{0.28\textwidth}{\includegraphics[width=0.28\textwidth]{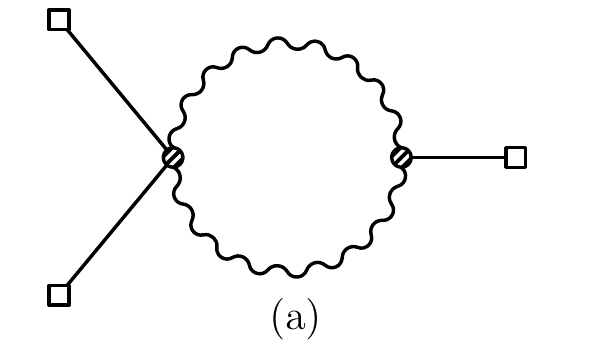}}~~+~~
  \parbox{0.28\textwidth}{\includegraphics[width=0.28\textwidth]{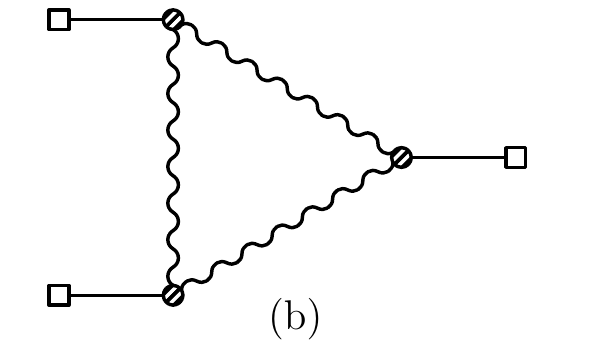}}
\end{equation}
Here we have used the diagrammatic representation introduced in \cite{Chen:2017ryl}. We first consider the Diagram (\ref{fd_vectorloop}a). Following the diagrammatic rules in \cite{Chen:2017ryl}, the amplitude corresponding to this diagram reads,
\begin{align}
  \text{Diag.\;(\ref{fd_vectorloop}a)}=&\FR{1}{2}\cdot\FR{\ii g^2}{2}\cdot\FR{\ii g^2 h_0}{2}\sum_{a,b=\pm}ab\int\FR{\di\tau_1}{|H\tau_1|^4} \FR{\di\tau_2}{|H\tau_2|^4}G_a(k_1;\tau_1)G_a(k_2;\tau_1)G_b(k_3;\tau_2)\n\\
  &~\times\int\FR{\di^3\mb q}{(2\pi)^3}\Big[D_{\mu\nu}(q,\tau_1,\tau_2)\Big]_{ab}\Big[D^{\mu\nu}(|\mb k_3-\mb q|,\tau_1,\tau_2)\Big]_{ab},
\end{align}
where $a$ and $b$ are SK contour indices, $G$ and $D_{\mu\nu}$ represent the propagators of the Higgs fluctuation $\de h$ and the $W$ boson. The Higgs is light and we can use approximately the massless propagator for $G$,
\begin{align}
  G_\pm(k;\tau)=\FR{H^2}{2k^3}(1\pm\ii k\tau)e^{\mp\ii k\tau}.
\end{align} 
The explicit expression for $D$ is collected in the Appendix but we will not use it at this point.
Here we have written down the expression for one $W$ boson. To take account of two charge eigenstates $W^\pm$, we need to multiply the expression by 2, and for the $Z$ boson, we need to replace the couplings as $g^2\to g^2/c_W^2$ and also replace the mass implicit in the propagator $D$.

The loop integral can be evaluated either directly or using the real-space representation. The latter was elaborated in \cite{Chen:2016hrz}. Here we shall quote the main steps and refer readers to \cite{Chen:2016hrz} for more details. The basic idea is that the loop integral can be written as
\begin{equation}
  \mathcal{I}_A(k;\tau_1,\tau_2)=\int\di^3X\,e^{-\ii \mb k\cdot \mb X}\la A_\mu(x) A^\mu(x) A_\nu(x')A^\nu(x')\ra,
\end{equation}
where $x=(\tau_1,\mb x)$, $x'=(\tau_2,\mb x')$ and $\mb X=\mb x-\mb x'$. The oscillatory signals in the squeezed bispectrum appear when $x$ and $x'$ are far apart, where the loop integral $\mathcal{I}_{A}$ develops non-analytic dependence on the spacetime coordinates as non-integer power of $\tau_1\tau_2/X^2$. We will call this part of the integral the nonlocal part, and it is free of UV divergence. The nonlocal part of the correlator $\la A^2(x)A^2(x') \ra$ was calculated in \cite{Chen:2016hrz}. The result in \cite{Chen:2016hrz} contains an incorrect factor whose effect is subdominant. After correcting this, we find the result as
\begin{align}
  \la A^2(x)A^2(x')\ra=\FR{3H^8}{64\pi^5m_A^4}\bigg[(1+2\mu_1)^2\Gamma^2(\mu_1)\Gamma^2(\FR{5}{2}-\mu_1)\Big(\FR{\tau\tau'}{X^2}\Big)^{3-2\mu_1}+\text{c.c.}\bigg],
\end{align}
where $m_A$ is the mass of the gauge boson and $\mu_1\equiv \sqrt{1/4-(m_A/H)^2}$. The above expression assumes that $\mu_1$ is complex, namely $m_A>H/2$. In the opposite case where $m_A<H/2$ and $\mu_1$ is real, one should drop the ``c.c.'' term. All expressions below should be understood in the same way. 

To calculate the loop integral $\mathcal{I}_A$ from the above correlator, we use the integral,
\begin{align}
  \int\di^3X e^{-\ii \mb k\cdot \mb X} X^{-2\ga}=\FR{4\pi}{k^{3-2\ga}}\Gamma(2-2\ga)\sin(\pi\ga).
\end{align}
Then it is straightforward to get
\begin{align}
\label{IAresult}
  \mathcal{I}_A(k;\tau_1,\tau_2)=\FR{3H^8}{16\pi^4m_A^4k^3}(1+2\mu_1)^2\Gamma^2(\mu_1)\Gamma^2(\FR{5}{2}-\mu_1)\Gamma(-4+4\mu_1)\sin(2\pi\mu_1)(k^2\tau_1\tau_2)^{3-2\mu_1}.
\end{align} 
Therefore we finally have
\begin{align}
 \text{Diag.\;(\ref{fd_vectorloop}a)}=&~\FR{3g^4H^6h_0}{1024\pi^4m_A^4 k_1^6}\bigg[\mu_1^{-2}(1+2\mu_1)^2(2-\mu_1)\sin^3(\pi\mu_1)\cos(\pi\mu_1)\n\\
  &~\times\Gamma^2(2-2\mu_1)\Gamma^2(\FR{5}{2}-\mu_1)\Gamma^2(\mu_1)\Gamma(-4+4\mu_1)\Big(\FR{k_3}{2k_1}\Big)^{-2\mu_1}+\text{c.c.}\bigg].
\end{align} 

The right diagram of (\ref{fd_vectorloop}) is more complicated. We shall make some simplifying assumptions. The hard internal line contributes mostly at the resonant point $|2k\tau|\simeq m/H$ and therefore we shall evaluate the propagator at this point without keeping all momentum and time dependence. Since the propagator at sub-horizon scales is $\order{H}$, and the additional time integral over $\tau_2$ would also contribute a factor of $H^{-1}$, we see that the net effect will be multiplying the result of the first diagram by a factor of $2g^2h_0^2/H^2$, where the factor of 2 is because (\ref{fd_vectorloop}a) contains a symmetry factor $1/2$ but (\ref{fd_vectorloop}b) does not. So, using (\ref{shape}), the shape function from both diagrams can be written as
\begin{align}
   \mathcal{S}_A(k_1,k_2,k_3;g,m_A)=&~R_h^3\FR{3g^4H^3h_0}{512\pi^5m_A^4}\Big(1+\FR{2g^2h_0^2}{H^2}\Big)\bigg[\mu_1^{-2}(1+2\mu_1)^2(2-\mu_1)\sin^3(\pi\mu_1)\cos(\pi\mu_1)\n\\
  &~\times\Gamma^2(2-2\mu_1)\Gamma^2(\FR{5}{2}-\mu_1)\Gamma^2(\mu_1)\Gamma(-4+4\mu_1)\Big(\FR{k_3}{2k_1}\Big)^{2-2\mu_1}+\text{c.c.}\bigg].
\end{align} 
Applying the above result to $W$ and $Z$ bosons, we get the corresponding shape functions,
\begin{align}
  &\mathcal{S}_W=2\mathcal{S}_A(k_1,k_2,k_3;g,m_W),
  &&\mathcal{S}_Z=\mathcal{S}_A(k_1,k_2,k_3;g/c_W,m_Z).
\end{align}

\paragraph{Top quark signal.} Next we consider the top quark's contribution to $\la\de h^3\ra$ at 1-loop level, from the following diagram.
\begin{equation}
\label{fd_fermionloop}
  \parbox{0.28\textwidth}{\includegraphics[width=0.28\textwidth]{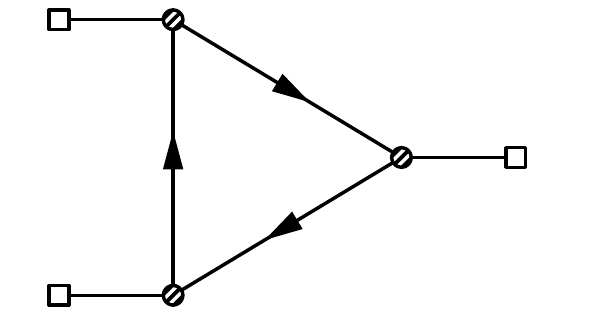}}
\end{equation}
Again, there are only two soft internal lines contributing to the nonlocal part of the diagram, and we will approximate the hard internal line by a factor of $1/H$. Then the diagram can be written as
\begin{align}
\label{toploop}
  \text{Diag.\;(\ref{fd_fermionloop})}=& ~2\cdot3\cdot\FR{\ii y_t^2}{2H}\cdot\FR{\ii y_t}{\sqrt 2}\sum_{a,b=\pm}ab\int\FR{\di\tau_1}{|H\tau_1|^4}\FR{\di\tau_2}{|H\tau_2|^4}G_a(k_1;\tau_1)G_b(k_2;\tau_2)G_c(k_3;\tau_2)\mathcal{I}_\psi(k_3;\tau_1,\tau_2).
\end{align}
Similar to our treatment for the gauge boson loop, here we have written the loop integral as the expectation value constructed from a 2-component Weyl spinor $\psi$,
\begin{equation}
  \mathcal{I}_\psi(k;\tau_1,\tau_2)=\int\di^3X\,e^{-\ii \mb k\cdot \mb X}\la \bar \psi \psi(x)\bar \psi\psi(x')\ra,
\end{equation}
The factor $2$ in front of (\ref{toploop}) takes account of the fact that a top quark can be written as two independent Weyl spinors, and the factor $3$ counts 3 colors. The nonlocal part of the correlator has been calculated in \cite{Chen:2018xck} and the result is,
\begin{align}
  \la \bar \psi \psi(x)\bar \psi\psi(x')\ra=-\FR{3H^6}{4\pi^5}(1-2\ii\wt m_t)\Gamma^2(2-\ii\wt m_t)\Gamma^2(-\FR{1}{2}+\ii\wt m_t)\Big(\FR{\tau\tau'}{X^2}\Big)^{4-2\ii\wt m}+\text{c.c.}.
\end{align}
From this we can find the following result for the nonlocal part of Diag.\;(\ref{fd_fermionloop}), 
\begin{align}
  \text{Diag.\;(\ref{fd_fermionloop})}
  =&~\FR{9\ii y_t^3H^3}{\sqrt{2}\pi^4k_1^6}\bigg[\FR{5-2\ii\wt m_t}{(1-2\ii \wt m_t)^3}\cosh^3(\pi\wt m_t)\sinh(\pi\wt m_t)\n\\
  &~\times\Gamma^2(2-\ii\wt m_t)\Gamma^2(\FR{1}{2}+\ii \wt m_t)\Gamma^2(3-2\ii\wt m_t)\Gamma(-6+4\ii \wt m_t)\Big(\FR{k_3}{2k_1}\Big)^{1-2\ii \wt m_t}+\text{c.c.}\bigg],
\end{align}
where $\wt m_t=m_t/H$. Then the shape function of the top signal is
\begin{align}
  \mathcal{S}_\text{top}=&~R_h^3\FR{9\sqrt{2}\ii y_t^3H^3}{\pi^5 }\bigg[\FR{5-2\ii\wt m_t}{(1-2\ii \wt m_t)^3}\cosh^3(\pi\wt m_t)\sinh(\pi\wt m_t)\n\\
  &~\times\Gamma^2(2-\ii\wt m_t)\Gamma^2(\FR{1}{2}+\ii \wt m_t)\Gamma^2(3-2\ii\wt m_t)\Gamma(-6+4\ii \wt m_t)\Big(\FR{k_3}{2k_1}\Big)^{3-2\ii \wt m_t}+\text{c.c.}\bigg].
\end{align}

\paragraph{Summary.} Above we have calculated the oscillatory signals in the squeezed bispectrum from $W/Z$ boson and the top quark. In all cases the shape function $\mathcal{S}$ exhibits non-integer power dependence on the momentum ratio $k_3/k_1$, with the following form,
\begin{equation}
  \mathcal{S}=\bigg[\mathcal{C}(m)\Big(\FR{k_3}{k_1}\Big)^{\al+\ii\be}+\text{c.c.}\bigg]=2\big|\mathcal{C}(m)\big|\Big(\FR{k_3}{k_1}\Big)^{\al}\cos\Big(\be\log\FR{k_3}{k_1}+\de\Big).
\end{equation}
Here $\mathcal{C}(m)$ is a complex coefficient dependent on the loop particle's mass $m$ and also its coupling to the Higgs field. $\al $ and $\be$ are real numbers, and they also depend on the mass $m$. In the second equality, we spell out the oscillation explicitly, with the frequency $\be$ and the phase $\de$. We define the ``clock amplitude'' $f_\text{NL,clock}$ by $f_\text{NL,clock}=2|\mathcal{C}|$ which provides a dimensionless measure of the signal strength.  Compare this number with the observational constraint/limit can provides us a rough idea of the visibility of the signals. The name ``clock'' is from the fact that such signals can also be used as ``standard clocks'' measuring the expansion history of the primordial universe, although the signals is truly oscillatory only when $m$ is large enough ($m>\frac{3}{2}H,0,\frac{1}{2}H$ for spin 0,$\frac{1}{2}$,1 fields, respectively). 

\begin{figure}[tbph]
\centering 
\vcenteredhbox{\includegraphics[width=0.5\textwidth]{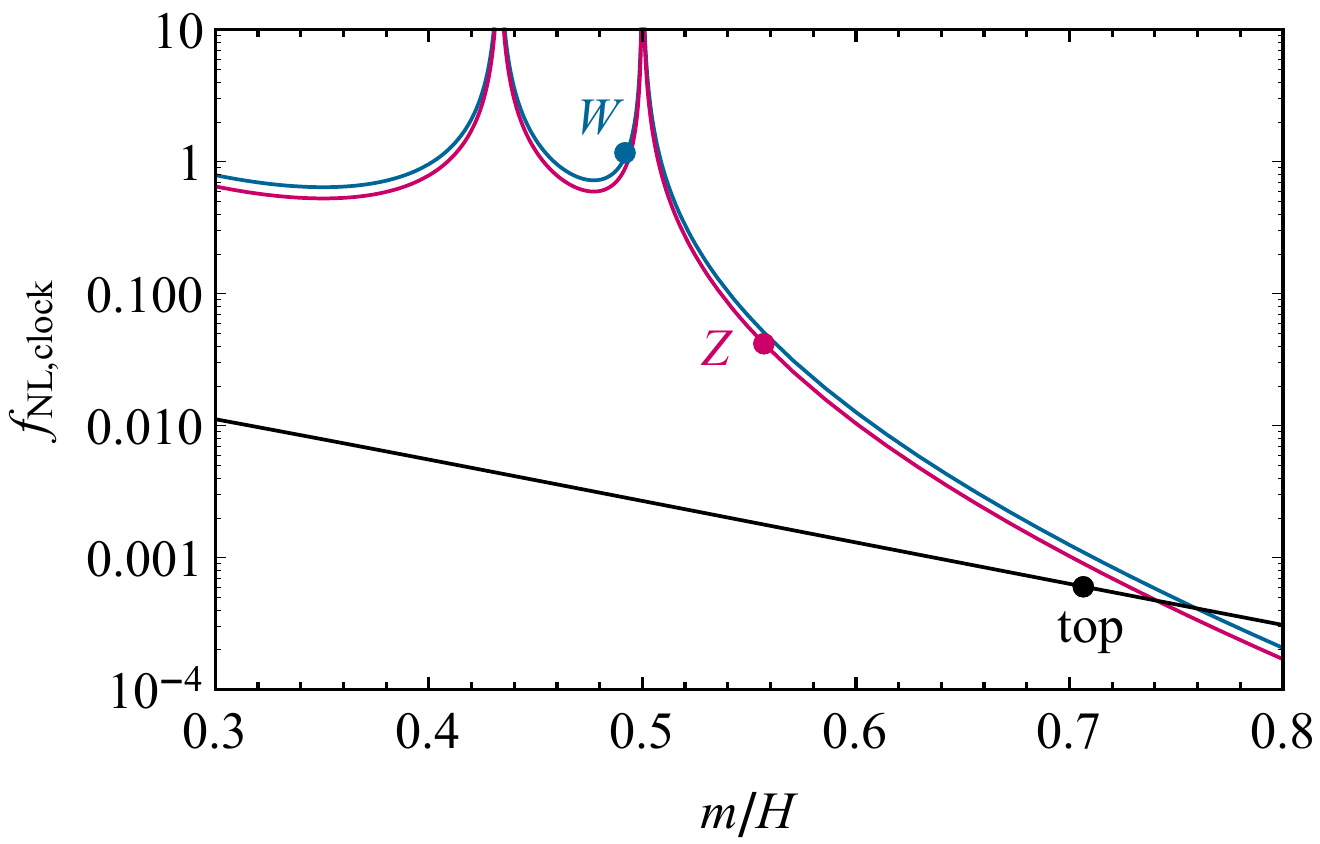}}
\caption{SM signals at CHC. The blue, purple, and black curves show the signals strength $f_\text{NL,clock}$ as functions of particles' mass for $W$, $Z$, and top quark, respectively. The choice of parameters are $R_h=0.14$, $\lam=0.01$, $g=0.6$, $y_t=1$ and $c_W=0.88$. Even larger non-Gaussianities are possible if smaller $\lambda$ is chosen.}
\label{Fig_signal}
\end{figure}

In the case of a detection of clock signals, in principle we can measure both the amplitude $f_\text{NL,clock}$ and the frequency $\be$. The frequency is directly related to the mass of the intermediate particle, while the amplitude depends on both the mass and the coupling to the external lines, namely the Higgs field in our case. Both of these quantities are calculable as elaborated above, and in Fig.~\ref{Fig_signal} we show the signals from $W/Z$ and top loops calculated above in the $(f_\text{NL,clock},m)$ plane. For illustration we take $R_h=0.14$, and for SM parameters we take approximately $\lam=0.01$, $g=0.6$, $y_t=1$, $c_W=0.88$. Of course these parameters also undergo renormalization group running from the electroweak scale where they are measured to the inflation scale. In general RG running introduces small $\order{1}$ corrections. But in the special case of Higgs self-coupling, RG  running could also introduce order-of-magnitude change. In fact, our choice of $\lam=0.01$ in Fig.~\ref{Fig_signal} is partially motivated by the fact that $\lam$ can be much smaller during inflation than its value at the electroweak scale. Of course if one sticks to the result of SM RG running, then $\lam$ will turn negative for high scale inflation. For our purpose of realizing CHC, we will assume either that the inflation scale is lower than the scale of Higgs instability or that new physics will change the RG running of Higgs self-coupling at high scales.

Given the fact that the clock amplitude $f_\text{NL,clock}$ is exponentially sensitive to the mass of the loop particle, and that our knowledge of particles' mass during inflation is not as precise, we also vary the masses of the internal particles in Fig.~\ref{Fig_signal} to illustrate the variations in the signal strength.  The two peaks at $m/H=\sqrt{3/16}$ and $m/H=1/2$ in Fig.~\ref{Fig_signal} in $W$ and $Z$ curves come from the factor $\Gamma(-4+4\mu_1)$ in the loop integral (\ref{IAresult}), which signals the failure of convergence of the loop integral. More careful treatment of the loop integral should remove this divergence. So, even we may expect enhancement of the signals near these two values of $m/H$, the two peaks themselves in Fig.~\ref{Fig_signal} should be taken with a grain of salt.

\subsection{Signals from beyond SM couplings}

Above we showed that the SM fields appear in the CHC signals at 1-loop if we assume that they couple to the Higgs field only through SM couplings. However, beyond-SM couplings can well be present during inflation. Some of these new couplings will lead to interesting new signals that do not exist within SM. In this subsection we consider one such example from Sec.~\ref{sec_higgsChannel}, namely the $\Lambda_{hi}$-term in (\ref{dim5Higgs}),
\begin{equation}
  \Delta\ld=\FR{g}{2c_W\Lambda_{hi}}(\pd_\mu\phi)Z^\mu h^2\To -\FR{g\dot\phi_0}{2c_W\Lambda_{hi}}a^{-2}(\tau)Z_0 (h_0+\de h)^2.
\end{equation}
We can have a tree-level diagram propagating $Z^0$ boson in the squeezed spectrum,
\begin{equation}
\label{fd_Ztree}
  \parbox{0.3\textwidth}{\includegraphics[width=0.3\textwidth]{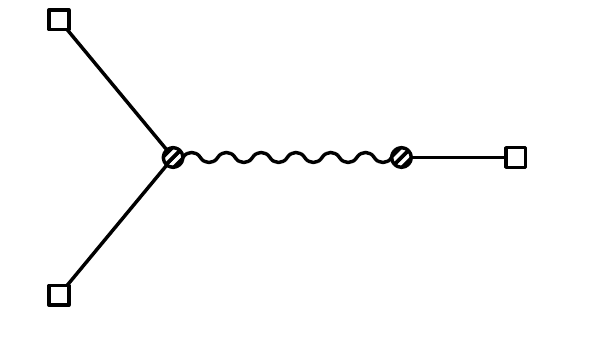}}
\end{equation}
We can simply estimate the magnitude of the oscillatory signals from the above diagram as,
\begin{equation}
  f_\text{NL}~\text{of Diag.~(\ref{fd_Ztree})}\sim  \FR{R_h^3P_\zeta^{-1/2}}{2\pi}\bigg(\FR{ \dot\phi_0}{ \Lambda_{hi}}\bigg)^2h_0.
\end{equation}
As before, we have taken $H=1$ in this estimate.
Given $m_Z\sim h_0\sim H$, the unitarity bound $\Lambda_{hi}>\dot\phi_0^{1/2}$, and the relation $\dot\phi_0^{1/2}> 60H $, we see that the magnitude of (\ref{fd_Ztree}) can be huge. With high scale inflation $H\sim 10^{13}\text{GeV}$, even $\Lambda_{hi}\sim \Mp$ will give $f_\text{NL}\sim \order{1}R_h^3$. A lower cutoff scale can give rise to clock signals much larger than that of the SM. In this manner the CHC may be sensitive to BSM physics. Now let us evaluate the diagram (\ref{fd_Ztree}) using diagrammatic rules,
\begin{align}
\text{Diag.~(\ref{fd_Ztree})}=  \Big(-\FR{\ii g\dot\phi_0}{c_W\Lambda_{hi}}\Big)^2h_0\sum_{a,b=\pm}ab\int\FR{\di \tau_1}{|H\tau_1|^3}\FR{\di \tau_2}{|H\tau_2|^3}G_a(k_1;\tau_1)G_a(k_2;\tau_1)G_b(k_3;\tau_1)\big[D_{00}(k_3;\tau_1,\tau_2)\big]_{ab}.
\end{align}
Here $D_{00}$ is the $00$-component of the propagator for $Z$. The $0$ component of polarization tensor is nonvanishing only for the longitudinal mode. Its polarization vector is
\begin{equation}
e_\mu^{(L)}(\mb k)=\bigg(-\FR{k}{m_Za},\sqrt{1+\Big(\FR{k}{m_Za}\Big)^2}\wh{\mb k}\bigg).
\end{equation} 
Therefore we have (see the appendix),
\begin{align}
  D_{00}(k_3;\tau_1,\tau_2)=\FR{2}{k}\FR{H^2}{m_Z^2}\bigg[\FR{\Gamma^2(-\mu_1)}{\pi}(k^2\tau_1\tau_2/4)^{\mu_1+3/2}+(\mu_1\to-\mu_1)\bigg].
\end{align}
The result is
\begin{align}
\mathcal{S}=R_h^3\FR{g^2\dot\phi_0^2h_0P_\zeta^{-1/2}}{2^{5/2}\pi^2c_W^2\Lambda_{hi}^2H m_Z^2}\bigg[\FR{(7-2\mu_1)(1+\sin\pi\mu_1)\Gamma^2(\mu_1)\Gamma^2(\frac{3}{2}-\mu_1)}{(1+2\mu_1)^2}\bigg(\FR{k_3}{4k_1}\bigg)^{3/2-\mu_1}+\text{c.c.}\bigg].
\end{align}
We note that the shape function does not depend on the angle between $\mb k_1$ and $\mb k_3$ as one may naively expect from an internal spin-1 line. This is because the $Z$ boson couple to the external Higgs background through its 0-component. So it behaves effectively like a scalar field.

\section{Discussions}
\label{sec_discussions}

The modulated reheating provides a simple way to generate primordial density fluctuations. In the context of the cosmological collider physics, we can say that the modulated reheating scenario turns inflation into a light scalar collider, where the light scalar is the field modulating the inflaton decay. 

In this paper we have shown that the SM Higgs boson can be a perfect modulating field. This Higgs-modulated reheating scenario thus provides us a Higgs collider working at the inflation scale and opens up new possibilities of studying Higgs physics at energies potentially far above any ground-based colliders. 

We have provided a simple realization of Higgs-modulated reheating where the inflaton decays into Higgs-portal scalars, which can also be a dark matter candidate. We have calculated the SM signals at the CHC in this scenario. We have seen that the oscillatory/non-analytic scaling signals in the squeezed bispectrum can be naturally large.

As a demonstration of using CHC, we further studied signals of SM fields in the squeezed bispectrum, including the massive gauge boson loop and the top loop, which could be the leading signals of SM. In preferable scenarios one could observe them simultaneously.

We see that the non-Gaussianities produced in the Higgs-modulated reheating scenario are generally quite large. In particular, the large local-shape non-Gaussianity limits the parameter range of CHC. As a result, the density perturbation cannot be generated all by the Higgs (unless the Higgs self-coupling is very small). Consequently, the CHC signals are suppressed by a factor of $R_h^3$. But as we see from Sec.~\ref{sec_CHCsignal}, the size of CHC signals are exponentially sensitive to the mass of intermediate particles and they can be enhanced a lot by changing the mass by a small amount. So, the CHC signal in this case can still be observably large.

A far more interesting possibility is to study the signals of new physics. The CHC provides an unmatched energy for Higgs collisions, and one can imagine to study any new physics scenario associated with the Higgs boson at high energies up to $H_\text{inf}\sim 10^{13}$GeV. Some simple examples include a Higgs-portal scalar, and BSM Higgs-$Z$ couplings, which we briefly considered. We leave more studies in this respect for future works.

In a broader sense, CHC can be considered as an example of a class of cosmological isocurvature colliders. The isocurvature perturbation may have converted to curvature perturbation through curvaton \cite{Enqvist:2001zp, Lyth:2001nq, Moroi:2001ct}, modulated reheating or multi-brid \cite{Sasaki:2008uc, Huang:2009vk} mechanisms. Alternatively, the isocurvature perturbation may survive and be directly observed for very light fields such as axions. These isocurvature fluctuations may carry the information of mass and spin of the heavy particles during inflation. We hope to study these possibilities in the future.

\paragraph{Acknowledgment} We thank Xingang Chen for discussions. YW thanks Dong-Gang Wang for discussions. ZZX thanks Prateek Agrawal, Junwu Huang, Hayden Lee, Davide Racco, and Liantao Wang for discussions. SL and YW are supported in part by
ECS Grant 26300316 and GRF Grant 16301917 and 16304418 from the Research Grants Council of Hong Kong. YW thanks University of Warsaw for hospitality where part of this work was done. ZZX thanks TD Lee Institute for their hospitality when this work was in progress.

\begin{appendix}

\section{SK Propagators of a Spin-1 Field}

In this appendix we work out the mode function and the late-time limit of the propagator for a spin-1 massive gauge field. We will include a chemical potential for the gauge boson arising from the coupling $\phi F\wt F$ where $\phi$ is the inflaton. 

We begin with the following Lagrangian,
\begin{equation}
  \ld = \sqrt{-g}\bigg[-\FR{1}{4}F_{\mu\nu}F^{\mu\nu}-\FR{1}{2}m^2A_\mu A^\mu\bigg]-\FR{1}{4}\lam\phi F_{\mu\nu}\wt F^{\mu\nu},
\end{equation}
where $\wt F^{\mu\nu}=\frac{1}{2}\ep^{\mu\nu\rh\si}F_{\rh\si}$, $\ep^{0123}=1$.
The equation of motion of $A_\mu=(A_0,\mb A)$ can be derived, after imposing the gauge condition $\pd_\mu (\sqrt{-g}A^\mu)=0$, to be
\begin{equation}
  \mb A''-\nabla^2\mb A + a^2m^2\mb A - \lam a\dot\phi_0\nabla\times\mb A=0
\end{equation}
Or, in the momentum space,
\begin{equation}
  \mb A''+k^2\mb A+a^2m^2\mb A-\ii a\lam\dot\phi_0\mb k\times \mb A=0.
\end{equation}
Let $\mb k=(0,0,k)$ be in the 3-direction. Then we can redefine $A_\pm = \frac{1}{\sqrt 2} (A_1\pm \ii A_2)$ and $A_L=A_3$. Then the equation of motion becomes
\begin{align}
  & A_\pm''+(k^2 +a^2m^2 \pm 2a\mu k )A_\pm=0, 
  &&A_L''+(k^2+a^2m^2)A_L=0,
\end{align}
where $\mu\equiv \lam\dot\phi_0$.
The solutions with positive-frequency initial condition are
\begin{align}
  &A_\pm = \FR{e^{\mp\pi\mu/2H}}{\sqrt{2k}} \text{W}_{\pm\ka,\nu}(2\ii k\tau),
  &&A_L= \FR{\sqrt{\pi}}{2}e^{\ii\pi\nu/2}\sqrt{-\tau}\text{H}_{\nu}^{(1)}(-k\tau),
\end{align}
where $\text{W}_{\ka,\nu}$ is one of the Whittaker functions and $\text{H}_{\nu}^{(1)}$ is H\"ankel function of first kind. The indices are $\ka\equiv \ii\mu/H$, $\nu\equiv \sqrt{1/4-(m/H)^2}$. The normalization is determined by the canonical commutation relation $[A_i(\mb x),A_j'(\mb y)]=\ii\de_{ij}\de^{(3)}(\mb x-\mb y)$. The late-time $(\tau\to0)$ behavior of these modes are
\begin{align}
  &A_\pm\simeq \FR{e^{-\ii \pi(1/4+\nu/2)}}{\sqrt{2k}}\FR{e^{\mp \pi\mu/2}\Gamma(-2\nu)}{\Gamma(\frac{1}{2}\mp\ii\mu-\nu)} (-2k\tau)^{\nu+1/2}+(\nu\to-\nu),\\
  &A_L\simeq\FR{e^{-\ii\pi(1/2-\nu/2)}}{\sqrt{2k}}\FR{\Gamma(-\nu)}{\sqrt{\pi}}(-k\tau/2)^{\nu+1/2}+(\nu\to-\nu).
\end{align}
The non-local parts of the propagators are
\begin{align}
  D_{\mu\nu}(\mb k;\tau_1,\tau_2)=&\sum_{\al}e_\mu^\al(\mb k)e_\nu^\al(\mb k)D^{(\al)}(k;\tau_1,\tau_2),\\
  D^{(\pm)}(k;\tau_1,\tau_2)=&~\FR{e^{\mp\pi\mu}}{2k}\bigg[\FR{\Gamma^2(-2\nu)}{\Gamma(\frac{1}{2}+\ii\mu-\nu)\Gamma(\frac{1}{2}-\ii\mu-\nu)}(4k^2\tau_1\tau_2)^{\nu+1/2}+(\nu\to-\nu)\bigg],\n\\
  D^{(L)}(k;\tau_1,\tau_2)=&~\FR{1}{2k}\bigg[\FR{\Gamma^2(-\nu)}{\pi}(k^2\tau_1\tau_2/4)^{\nu+1/2}+(\nu\to-\nu)\bigg].
\end{align}
We see that only the transverse parts are affected by the chemical potential. The chemical potential will bring exponential enhancement/suppression to the two transverse polarizations, and not-too-large back reaction will put stringent bound on the cutoff scale $\Lambda$ as discussed in Sec.~\ref{sec_fer_gauge_channel}.

\end{appendix}

\providecommand{\href}[2]{#2}\begingroup\raggedright\endgroup

\end{document}